\documentclass[iop]{emulateapj}

\usepackage{amsmath}
\usepackage{amssymb}
\usepackage{graphicx}

\shorttitle{Oceans on Super-Earths}
\shortauthors{Schaefer \& Sasselov}
\begin{document}
\title{The persistence of oceans on Earth-like planets: insights from the deep-water cycle}
\author{Laura Schaefer and Dimitar Sasselov}
\affil{Harvard-Smithsonian Center for Astrophysics, 60 Garden St., Cambridge, MA 02138}
\email{lschaefer@cfa.harvard.edu}

\begin{abstract}
In this paper we present a series of models for the deep water cycle on super-Earths experiencing plate tectonics. The deep water cycle can be modeled through parameterized convection models coupled with a volatile recycling model. The convection of the silicate mantle is linked to the volatile cycle through the water-dependent viscosity. Important differences in surface water content are found for different parameterizations of convection. Surface oceans are smaller and more persistent for single layer convection, rather than convection by boundary layer instability. Smaller planets have initially larger oceans but also return that water to the mantle more rapidly than larger planets. Super-Earths may therefore be less habitable in their early years than smaller planets, but their habitability (assuming stable surface conditions), will persist much longer.
\end{abstract}
\maketitle

\section{Introduction}

In the post-Kepler era, the number of known planets of Earth and super-Earth size continues to grow. The TESS mission, which is a planned survey of the brightest and closest stars, is expected to detect thousands of small planets \citep{Ricker15}. The TESS planets, by virtue of their closeness and the brightness of their stars, will be much easier to follow up with ground-based photometric and spectroscopic techniques than the planets of similar size detected by Kepler. Currently, the mini-Neptune GJ 1214 is the only planet in the super-Earth mass range ($1-10 M_{\oplus}$) for which atmospheric measurements have been made \citep[e.g.,][]{W14,Fr13,B12}. With future telescope resources such as GMT and JWST, we will be able to characterize even more super-Earths, some of which may be in the habitable zones (HZ) of their stars. The classical habitable zone is defined as the orbital region in which an Earth-like planet can maintain liquid water on its surface given a variety of atmospheric compositions. However, the Earth\textquoteright s atmosphere, although it permits surface water to remain liquid, does not control the supply of water on the surface. For that, we must look to the mantle. In extending this to super-Earths, it is presently unclear what the effect of mass may be in controlling the supply of water at the surface. \\
\indent Water on the Earth is found in two primary reservoirs: (1) the surface oceans and (2) the silicate mantle. Water, either in molecular form or as an H$^{+}$ or OH$^{-}$ group, can dissolve in silicate minerals. This is true even of the so-called nominally anhydrous minerals (NAMs), which do not have hydrogen in their chemical formula. Furthermore, the solubility of water in silicate minerals increases with pressure \citep{Kohl96,H05}.\\
 \indent Geochemical studies of the materials that make up Earth\textquoteright s mantle have long suggested that it may hold anywhere from 1 to 10 times as much water as is found in the surface oceans \citep[e.g., Table 1 of][]{B01}, although more recent work seems to discount larger values in favor of values ranging from 0.5 to 2.5 ocean masses (OMs) of water \citep{HK12}. The presence of water in these minerals strongly influences their material properties, such as melting temperatures, rheology, phase changes, electrical conductivity, etc. \citep[see e.g.,][]{HK96,K90,LO07}. Studies of the effect of water on silicate rheology show that the presence of even minor amounts of water can reduce the viscosity by several orders of magnitude \citep{KW93}. This can have a significant effect on the material flow of the mantle. \\
\indent In fact, it has been shown that the abundance of water on the Earth\textquoteright s surface is controlled by the deep water/silicate cycle, which is tied to plate tectonics. The abundance of water at the surface is a balance between the rate of return via volcanic outgassing from the mantle at mid-ocean ridges (MORs) and the rate of loss to the mantle through so-called ingassing, which is the return of water into the deep mantle by the sinking, or subducting, of water-rich oceanic seafloor. Much of this water is released immediately back to the surface through shallow, water-induced volcanism. However, a small but significant fraction of the water contained in the subducting oceanic slabs can be transported to deeper levels of the mantle. For a detailed review of the exchange mechanisms between the surface and mantle, see \cite{H06}.\\
\indent The deep water cycle on Earth has been studied through the use of parameterized convection models incorporating a water-dependent viscosity \citep[e.g.][]{MS89,B01,Sandu11,Cr11}. Parameterized convection models are simplified 1D models using a parameterization derived from more complicated and expensive 2D and 3D models of convection for different systems (e.g. spherical vs. plane-parallel, heated from within vs. below, etc.). The parameterized models rely primarily on two dimensionless parameters: the Rayleigh number, which describes whether a system is unstable to convection, and the Nusselt number, which compares the convective and conductive heat flows. The Rayleigh number depends on the viscosity of the system, which itself is dependent on temperature, pressure and water fugacity. The abundance of water in the mantle, which helps determine the viscosity, evolves along with the mantle temperature. \\
\indent Here we will use a parameterized convection model to study the deep water cycles of super-Earths. In the current era of exoplanet studies, we are still searching for Earth-like planets in Earth-like orbits around Sun-like stars, but what we have found and what we can soon characterize, are super-Earths ($\sim 1-2 R_{\oplus}, \sim 1-10 M_{\oplus}$). Although super-Earths may have cool enough surfaces to retain liquid water, tectonics and mantle convection plays an important role in controlling its abundance at the surface. The question of whether these planets will have plate tectonics has been discussed previously in the literature \citep[e.g.][]{Val07,ONL07, K10b, NB13,SB14}, although the issue has not been settled. Here, we assume plate tectonics as a starting point in order to apply a similar water cycle model. The question this work then addresses is whether the different pressure regimes inside super-Earths affect the deep water cycle, and whether surface oceans are tectonically sustainable on these planets.  This has important implications for the potential habitability of super-Earths. \\
\indent In this paper, we address these questions using a parameterized thermal evolution model coupled with a water cycle model. We give a full description of the model in Section 2. In Section 3, we describe results for models which have either single layer convection or boundary layer convection for a number of super-Earth models. We also describe the response of the models to reasonable variations in the parameter values. Section 4 discusses implications for the persistence of oceans on super-Earths. 

\section{Methods}

Mantle convection in the terrestrial planets can be controlled either by the stability of the whole mantle layer (single layer convection, e.g. \cite{S01}) or by the stability of two boundary layers: a cold boundary layer at the surface and a hot boundary layer at the interface of the silicate mantle with the metallic core \citep[see e.g.][]{TS02}. Heat from either secular cooling or decay of radioactive materials (or both) is transferred by conduction through the boundary layers. The interior region is convective and thus approximately adiabatic. See Figure 1 for a schematic representation of the thermal profile. Single layer convection models typically neglect the heat flux from the core, which is small, for simplicity and are entirely heated from within by radioactive decay. Numerical simulations are used to determine scaling laws relating the heat flux and mean temperature to the degree of convection. These simulations have been done for a variety of different boundary conditions (free slip, no slip, etc.), and for different material properties (isoviscous, highly temperature-dependent viscosity, etc.) \citep[e.g.,][]{HI96,DS00,D05}. Choosing the proper scaling relations is therefore important for a successful model. \\
	\begin{figure}
	\plotone{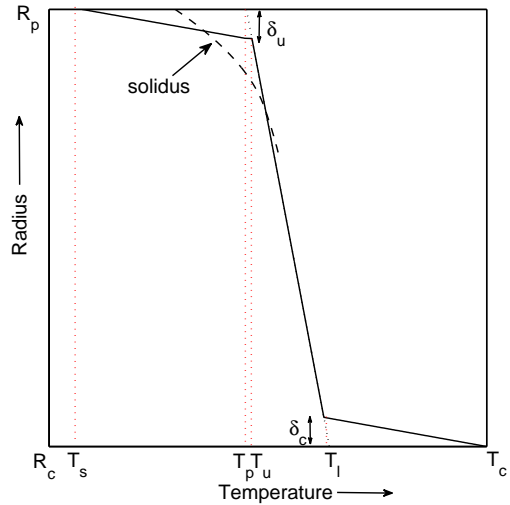}
	\caption{Schematic of temperature-depth profile. See text for details.}
	\end{figure}
\indent The parameter which is of the most importance in determining the convective behavior is typically the characteristic viscosity. Discussion in the literature is often contradictory on where to define the characteristic viscosity: as an average value \citep[see e.g.][]{MS89,TM92,Sandu11}, or at the base of a boundary layer \citep[see e.g.][]{DS00,D05,Vlada12}. Here we will try both types of models, starting with the single-layer models typically used with volatile evolution models, which use average viscosities. We will then look at a boundary layer model. We explore both models here to see the effect on the behavior of water, and note that we are not attempting to reproduce the Earth, but explore two valid, but different, models.\\
\indent The water cycle model is coupled to the thermal model through the viscosity, which is dependent on the water abundance in the mantle \citep[see e.g.][]{MS89,Sandu11}. Models of the Earth have shown that the viscosity and mantle temperature create a feedback loop \citep{S01,Cr11}. When the mantle is warm, convection is vigorous and the mantle cools quickly. As the temperature drops, the viscosity increases, causing the convection to become sluggish. Sluggish convection means that less heat is removed from the mantle, causing it to heat up. The viscosity increases, and the cycle repeats. This cycle has been shown to be enhanced by the presence of water \citep{MS89,Cr11}. \cite{Cr11} describe how the water and temperature feedbacks interact for temperature and water-dependent viscosities. \cite{CA14} studied the effect of sea floor pressure on a steady state model of the deep water cycle, without considering the effect of planetary thermal evolution. \\
\indent In the following sections, we first describe selection of the super-Earth model parameters. We then describe the parameterized thermal model, followed by the water cycle model. Section 3 will discuss the results for these models. 

\subsection{Super-Earth Models}

We use the scaling relations of \cite{Val06} to calculate the planetary parameters for planets of 1, 2, 3, and 5 Earth masses ($1 M_{\oplus} = 5.97\times10^{24}$ kg). Larger planets are not considered here because \cite{NB13} find that the peak likelihood for plate tectonics occurs for planets between $1 - 5 M_{\oplus}$. The scaling laws of \cite{Val06} assume a constant core mass fraction of 0.3259. The planetary and core radii then scale by $R_{i} \sim R_{i,\oplus} (M_{p} / M_{\oplus})^{a_{i}}$, where $i  = c$ (core) or $p$ (planet), $a_{c}$ = 0.247, $a_{p}$ = 0.27, and values for the Earth are indicated by $\oplus$. The average mantle density $\langle \rho_{m} \rangle$ is calculated from the mantle mass and volume. The average gravitational acceleration $\langle g \rangle$ is found from $GM_{p} / R_{p}^{2}$. Values for these parameters are given in Table 1. We take a constant water mantle mass fraction of $1.4\times10^{-3}$ for the nominal models. This is equivalent to 4 ocean-masses (OM) of water for the Earth, where 1 OM is equal to $1.39\times10^{21}$ kg H$_{2}$O. We will later explore the effect of variable water on the results.

	\begin{deluxetable}{c c c c c }
	\tablecaption{Super-Earth Model Parameters.}
	\tablehead{\colhead{\parbox[t]{4em}{\centering Mass \\ ($M_{\oplus}$)}} & \colhead{\parbox[t]{4em}{\centering $R_{p}$\\ ($R_{\oplus}$) }} & \colhead{\parbox[t]{4em}{\centering $R_{core}$  \\($R_{\oplus}$)}} &\colhead{\parbox[t][][t]{5em}{\centering $\langle \rho_{m} \rangle$ \\(kg m$^{-3}$)}} & \colhead{\parbox[t]{4em}{\centering $\langle g \rangle$  \\(m s$^{-2}$)}}}
	\startdata
	1 & 1 & 0.547 & 4480 & 9.8\\
	2 & 1.21 & 0.649 & 4960 & 13.4\\
	3 & 1.34 & 0.717 & 5470 & 16.4\\
	5 & 1.54 & 0.814 & 5970 & 20.7
	\enddata
	\tablecomments{Core mass fraction is held fixed at 0.3259, and scaling relations from Valencia et al. (2006) are used to determine $R_{p}$ and $R_{core}$. See text for details.}
	\end{deluxetable}

\subsection{Thermal Evolution Model}

Following models of Earth\textquoteright s deep water cycle \citep[e.g.][]{MS89,S01,Sandu11}, we will first consider models heated only from within (i.e., heat flux $q_{c} = 0$). These models are considered single-layer convection because the whole mantle convects. There is a conductive upper thermal boundary layer that governs heat loss from the surface. For most of the lifetime of the Earth, such models have been shown to give good fits to the observed mantle viscosity and heat flux \citep{S01}. 
\indent The thermal evolution model requires solution of the mantle heat transfer equation:
	\begin{equation}
	\rho_m C_p V_m  \frac{d\langle T_m \rangle}{dt}=-A_s q_s+A_c q_c+V_m Q(t)
	\end{equation}
where $\rho_{m}$ is the mantle density, $C_{p}$ is the mantle heat capacity, $V_{m}$ is the mantle volume, $A_{s}$, $A_{c}$ are the surface area\textquoteright s of the planet and core, $q_{s}$ and $q_{c}$ are the conducted heat fluxes through the surface and core-mantle boundary (CMB), respectively, and $Q(t)$ is the heat produced by radioactive decay within the mantle. In this model, there is no heat flux from the core, so the second term on the right side vanishes. 
\indent The temperature modeled in any parameterized convection model is somewhat ambiguous. Here, we will follow the convention of \cite{MS89} and take the temperature to be the spherically-averaged mantle temperature. We can relate this averaged temperature to the temperature of the mantle adiabat extrapolated to the surface (i.e. the mantle potential temperature T$_{p}$) through the equation:
	\begin{equation}
	\langle T_{m} \rangle = \epsilon_{m}T_{p} = \frac{3}{R_{p}^{3} - R_{c}^{3}}\int_{R_{c}}^{R_{p}} r^{2}T(r)dr
	\end{equation}
See \cite{TM92} for a derivation of this equation. We approximate the adiabat as:
	\begin{equation}
	T(r) = T_{p} + T_{p}\frac{\alpha g}{c_{p}}\Delta r
	\end{equation}
Values of $\epsilon_{m}$ for the different planet masses are given in Table 5. \\
\indent In the mantle, heat is generated by decay of radioactive elements. The heat flux from radioactive decay is dominated by $^{238}$U, $^{235}$U, $^{232}$Th, and $^{40}$K. The heat produced is calculated from the equation:
	\begin{equation}
	Q(t)= \rho_m \sum C_i  H_i  exp⁡[\lambda_i (4.6 \times 10^9 - t)]
	\end{equation}
where $C_{i}$ is the mantle concentration of the element by mass, $H_{i}$ is the heat production per unit mass, and $\lambda_{i}$ is the decay constant. The decay constants, heat production rates, and abundances relative to total uranium are taken from \cite{TS02}. In this paper, we assume that all super-Earths have the same ratios of radioactive elements as the present day Earth. The nominal bulk silicate Earth (i.e., primitive mantle) contains $\sim$ 21 ppb U \citep{MS95}. \\
\indent Using boundary layer theory, the heat flux out of the mantle is given by:
	\begin{equation}
	q_m=k \frac{(T_u-T_s)}{\delta_u} 
	\end{equation}
where $k$ is the mantle conductivity, $\delta_{u}$ is the boundary layer thickness, and the T$_{u}$ is the temperature at the base of the boundary layer (see Figure 1). $T_{u}$ is calculated from $\langle T_{m} \rangle$ using the adiabatic temperature profile of the mantle. \\
\indent The thickness of the upper boundary layer is given by the global Rayleigh number:
	\begin{equation}
	\delta_u= Z \left( \frac{Ra_{cr}}{Ra} \right)^{\beta} =  \left( \frac{\kappa \eta(T, P) Ra_{crit}}{g \alpha \rho_m \Delta T} \right) ^{\beta}
	\end{equation}
where $Z$ is the thickness of the mantle, $Ra$ is the Rayleigh number of the whole mantle, $Ra_{crit}$ is the critical Rayleigh number for convective instability, $\kappa$ is the mantle thermal diffusivity, $\eta(T, P)$ is the characteristic mantle viscosity,  and $\alpha$ is the thermal expansivity. $\Delta T$ is the temperature drop across the mantle minus the adiabatic temperature change. We use the mantle viscosity calculated with $\langle T_{m} \rangle$ and $\langle P \rangle$, which is characteristic of the whole mantle layer. This choice dictates the behavior of our model, as will be discussed later. The viscosity parameterization is described in Section 2.4. \\
\indent Two other parameters are derived from the thermal model. The areal spreading rate is the rate at which new ocean crust is being created. \cite{MS89} parameterized the areal spreading rate using the current volume of the oceans and the present day heat flux, which are unconstrained for exoplanets. Instead, we follow \cite{Sandu11}, who relate the areal spreading rate to the convective velocity {$u_{c}$) and the length of the spreading centers ($L_{ridge}$) where ocean crust is created:
	\begin{equation}
	S=2L_{ridge} u_{c}
	\end{equation}
The convective velocity is determined by the convective layer overturn time from boundary layer theory \citep[ch. 8]{S01} using the equation:
	\begin{equation}
	u_c= \frac{5.38 \kappa (R_p-R_c)}{\delta_u^2}
	\end{equation}
We parameterize the length of the mid-ocean ridges as 1.5 times the planetary circumference. This parameterization is chosen to give the present day mid-ocean ridge (MOR) length on the Earth of $\sim$60,000 km. We describe results using smaller $L_{ridge}$ values in Section 4. 

\indent Values for the parameters used in the thermal evolution model are given in Table 2. The value of $Ra_{crit}$ for the mantle is taken from \cite{S01}. We use constant values for the heat capacity, thermal conductivity, thermal expansivity and thermal diffusivity, although these parameters are all known to be pressure-dependent. 

	\begin{deluxetable}{l l c c c}
	\tablecaption{Nominal Thermal Model Parameters.}
	\tablehead{\colhead{param.} & \colhead{units} & \colhead{Upper Mantle} & \colhead{Lower Mantle} & \colhead{Core}}
	\startdata
	$Ra_{crit}$ & & 1100 & $0.28Ra^{0.21}$ & --- \\
	$\alpha$ & $\times 10^{-5}$ K$^{-1}$ &  2 & 1 & ---\\
	$k$& W m$^{-1}$ K$^{-1}$ & 4.2 & 4.2 & --- \\
	$\kappa$ & m$^{2}$ s$^{-1}$& $10^{-6}$ & $10^{-6}$ & ---\\
	$C_{p}$ & J kg$^{-1}$ K$^{-1}$& 1200 & 1200 & 840\\
	$\rho$ & kg m$^{-3}$ & 3300 & $\langle \rho_{m} \rangle$\tablenotemark{a} & 8400
	\enddata
	\tablenotetext{a}{See Table 1}
	\end{deluxetable}

\subsection{Volatile Evolution Model}

	\begin{deluxetable}{l l c}
	\tablecaption{Water Cycle Parameters.}
	\tablehead{\colhead{param.} & \colhead{name} & \colhead{value}}
	\startdata
	$\chi_{r}$ & regassing efficiency & 0.03\\
	$\chi_{d}$ & degassing efficiency & 0.02 \\
	$f_{bas}$ & hydrated basalt fraction & 0.03\\
	$\rho_{bas}$ & basalt density (kg m$^{-3}$) & 3000\\
	$L_{ridge}$ & Mid-ocean ridge length & 1.5$\times (2 \pi R_{p})$\\
	$K$ & Solidus depression constant (K wt\%$^{-\gamma}$) & 43 \\
	$\gamma$ & Solidus depression coeffiient & 0.75\\
	$\theta$ & Melt fraction exponent & 1.5\\
	$D_{H_{2}O}$ & Silicate/melt partition coefficient & 0.01
	\enddata
	\end{deluxetable}

Volatile evolution models harken back to \cite{MS89}, repeated with variations by many others. The volatile evolution model involves calculation of outgassing and ingassing rates for water based on mantle melting and surface hydration. The rate of change of the water abundance in the mantle is given by combining the ingassing and outgassing rates: 
	\begin{equation}
	\frac{dM_{H_{2}O,m}}{dt} = r_{ingas} - r_{outgas}
	\end{equation}
This equation is solved simultaneously with the heat transfer equation at each time. In the simplest form of \cite{MS89}, the outgassing rate is parameterized as: 
	\begin{equation}
	r_{outgas} = \rho_{m,v}d_{m}S
	\end{equation}
where $\rho_{m,v}$ is the density of volatiles in the mantle,, $d_{m}$ is the depth of melting, and $S$ is the areal spreading rate of the mid-ocean ridges. In \cite{MS89}, $d_m$ is kept at a constant value of 100 km. The ingassing rate is parameterized as:
	\begin{equation}
	r_{ingas} = f_{bas}\rho_{bas}d_{bas}S\chi_{r}
	\end{equation}
where $f_{bas}$ is the mass fraction of volatiles in the hydrated basalt layer, $\rho_{bas}$ is the density of basalt, $d_{bas}$ is the average thickness of the basalt (held constant at 5 km) and $\chi_{r}$ is an efficiency factor reflecting the incomplete transport of the water in the hydrated basalt layer into the deep mantle. With all parameters except $S$ and $\rho_{m,v}$ held constant in equations (10) and (11), we found that all planets necessarily reached a steady state (i.e., $\frac{dM_{H_{2}O,m}}{dt} = 0$), where the mantle water abundance is given by setting $r_{ingas}$ equal to $r_{outgas}$, and solving:
	\begin{equation}
	\rho_{m,v} = \frac{M_{H_{2}O,m}}{V_{m}} = \frac{ f_{bas}\rho_{bas}d_{bas}S\chi_{r}}{d_{m}S}
	\end{equation}
	\begin{equation}
	M_{H_{2}O,m}= \frac{ f_{bas}\rho_{bas}d_{bas}\chi_{r}V_{m}}{d_{m}}
	\end{equation}
While a case may be made that the Earth is in a steady state, there is little reason to suppose that this is a necessary condition for all exoplanets experiencing plate tectonics. We therefore follow here the volatile evolution model of \cite{Sandu11}, described briefly below. In this model, the depths of melting and hydration are not held constant, but vary based on local temperatures. We found that in these models, steady-state was rarely achieved. Parameters used in the volatile evolution model are given in Table 3. The outgassing rate is given by: 
	\begin{equation}
	r_{outgas} = \rho_m \langle F_{melt} \rangle \langle X_{melt} \rangle D_{melt} S \chi_d
	\end{equation}
where $\langle F_{melt} \rangle$ and $\langle X_{melt} \rangle$ are the average fraction of melting and average abundance of water in the melt over the melt layer thickness, $D_{melt}$, $S$ is the areal spreading rate and $\chi_{d}$ is the degassing efficiency, which accounts for incomplete transport of water to the surface. Whereas \cite{MS89} used a constant value for the melt layer thickness, \cite{Sandu11} used the mantle thermal profile and the peridotite solidus to determine the melt layer thickness. The thermal profile used is composed of the conductive upper thermal boundary layer and the upper mantle adiabat, and the intersection of this profile with the hydrated solidus curve for peridotite determines where melt forms (see Fig. 1). Water dissolved in silicates lowers their solidus (the temperature at which partial melting begins), and the water partitions preferentially into the melt. Using the parameterization of \cite{Katz03} for wet melting, the solidus depression is given by:
	\begin{equation}
	T_{sol,wet}=T_{sol,dry} - \Delta T_{H_{2}O} = T_{sol,dry} - K X_{melt}^{\gamma}	
	\end{equation}
where $K$ and $\gamma$ are empirically determined constants for peridotite (see Table 3). The melt fraction and water abundance in the melt are determined where the mantle thermal profile is above the wet solidus temperature and are given by:
	\begin{equation}
	F_{melt} = \left[ \frac{T - T_{sol,wet}}{T_{liq,dry} -T_{sol,dry}} \right]^{\theta}
	\end{equation}
	\begin{equation}
	X_{melt}=\frac{X_{H_{2}O,m}}{D_{H_{2}O}+F_{melt} (1- D_{H_{2}O})}
	\end{equation}
where $X_{H_{2}O,m}$ is the water mass fraction in the mantle, $D_{H_{2}O}$ is the silicate/melt partition coefficient, and $\theta$ is an empirically determined exponent. The dry liquidus and solidus equations are taken from \cite{ZH94} and \cite{H09}, respectively. The melt fraction and water fraction are averaged over the melt zone thickness at each timestep for use in equation (9). \\
\indent The return of water from the surface back into the mantle through ocean plate subduction is described by:
	\begin{equation}
	r_{ingas} = f_{bas}\rho_{bas} D_{hydr} S \chi_{r}
	\end{equation}
where $f_{bas}$ is the mass fraction of water in a hydrated basalt layer of thickness $D_{hydr}$, $\rho_{bas}$ is the density of basalt, and $\chi_{r}$ is the regassing efficiency, which accounts for imperfect return of water to the mantle. The hydrated layer thickness is measured from the surface, down to the depth at which the temperature reaches the stability boundary of serpentinite (i.e., serpentine is not stable at lower depths). The upper thermal boundary layer has a linear conductive temperature profile (see eq. 5), where the temperature change with depth is $(T_{u} - T_{s}) / \delta_{u}$. The depth to the serpentine stability temperature is therefore given by:
	\begin{equation}
	D_{hydr}= \delta_{u}\frac{T_{serp}-T_{s}}{T_{u} - T_{s}} = k \frac{T_{serp}-T_{s}}{q_m}
	\end{equation}
where $T_{serp}$ is the highest temperature of serpentine stability at pressures below $\sim 3$ GPa and is $\sim 973$ K \citep{UT95}. The hydrated layer is necessarily smaller than the thermal boundary layer, and is further restricted to hold no more water than is present at the surface in a given instant in order to maintain water mass-balance. We use the values for $\chi_{r}$ and $\chi_{d}$ determined by \cite{Sandu11}.  \\
\indent A final word about parameterized volatile evolution models: As noted by a reviewer, the equations described above assume that water transported into the mantle is instantaneously transported throughout the mantle and available for outgassing. In the real world, of course, the mantle is not homogeneous and the spreading zone melt centers will likely become dehydrated before they can be replenished by advection from subducted water. Therefore outgassing rates calculated here are upper limits on the true values.  

\subsection{Viscosity}

Water dissolved in silicate minerals such as olivine reduces the mineral\textquoteright s strength \citep[e.g.,][]{CP81}, and therefore its viscosity. In experimental literature, the water dependence of the viscosity is parameterized via the water fugacity $f_{H_{2}O}$, which depends on temperature and pressure. The rheological or constitutive law relating stress $\tau$ and strain rate $\dot{\epsilon}$ for olivine is then given by:
	\begin{equation}
	\dot{\epsilon} =A_{crp} \tau^n d^{-p}ln f_{H_{2}O}^r exp\left(-\frac{E_a+PV_a}{R_{gas} T}\right)
	\end{equation}
where n = 1 for diffusion creep, $A_{crp}$ is a material parameter, $d$ is the grain size in microns, $E_{a}$ is the activation energy, $V_{a}$ is the activation volume, and $R_{gas}$ is the ideal gas constant. This equation shows that the response to stress depends on the pressure and temperature, as well as the water fugacity ($f_{H_{2}O}$) in a non-linear way. The viscosity is then derived from the constitutive law:
	\begin{equation}
	\eta_{eff} = \frac{\tau}{2\dot {\epsilon}}
	\end{equation}
We combine the constant parameters $A_{crp}$ and the grain size into a normalization factor $eta_{0}$ to arrive at the effective viscosity. The form of the effective viscosity is then:
	\begin{equation}
	\begin{split}
	\eta_{eff}= \eta_0 f_{H_{2}O}^{-r} exp\left[ \frac{E_a}{R_{gas}}  (\frac{1}{T}-\frac{1}{T_{ref}} ) + \right. \\ \left. \frac{1}{R_{gas}}  (\frac{PV_a}{T}-\frac{P_{ref} V_a}{T_{ref}} )\right]
	\end{split}
	\end{equation}
We normalize so that  $\eta$($T_{ref}$ = 1600, $X_{H_{2}O}$= 500 ppm, $P_{ref}$ = 0)= $10^{21}$ Pa s, which gives reasonable viscosities for the Earth. Parameters are given in Table 4, and are taken primarily from \cite{HK03} for wet diffusion in olivine. These authors measured values of $r$ of $\sim 0.7-1.2$ on polycrystalline olivine samples. Recent work on Si diffusion in olivine and single-crystal deformation experiments suggests that the dependence on water abundance may be much lower ($r = 0 - 0.33$) \citep{F13,G13}. However, more work needs to be done to reconcile these experiments with the polycrystalline experiments. In this paper we use results from the earlier studies for the nominal models, but we also explore the effect of different $r$ values on our findings.

	\begin{deluxetable}{lcc}
	\tablecaption{Viscosity Parameters.}
	\tablehead{\colhead{parameter} & \colhead{Olivine (wet)}& \colhead{Perovskite(dry)}}
	\startdata
	$\eta_{0}$ (Pa s) & $1.08 \times 10^{25}$ & $1 \times 10^{21}$\\
	$r$ & 1.0 & ...\\
	$E_{a}$ (kJ mole$^{-1}$) & 335 & 300\\
	$V_{a}$ (cm$^{3}$ mole$^{-1}$) & 4.0 & 2.5\\
	$R_{gas}$ (J mole$^{-1}$ K$^{-1}$) & \multicolumn{2}{c}{8.314}
	\enddata
	\end{deluxetable}

\indent To convert from mass fraction water in the mantle to fugacity $f_{H_{2}O}$, we use the formulation of \cite{Li08}, which is an empirical relationship between the water fugacity and the concentration of water in olivine ($C_{OH}$, atomic H/$10^{6}$ Si): 
	\begin{equation}
	ln f_{H_{2}O} = c_0+c_1 lnC_{OH}+c_2 ln^2 C_{OH} + c_3 ln^3 C_{OH}
	\end{equation}
where $c_{0} = -7.9859, c_{1} = 4.3559, c_{2} = -0.5742$, and $c_{3} = 0.0227$. This relationship is derived from  olivine solubility data between $\sim 1373 - 1600$ K and is only strictly valid within that temperature range. However, the relationship varies only marginally with temperature for a wide range of $C_{OH}$ and $f_{H_{2}O}$, so we apply it to the whole temperature range considered here for lack of other available data. It is straightforward to convert from the mantle water mass fraction to the concentration in $C_{OH}$ using the molecular weights of water and olivine.  \\

	\begin{deluxetable}{c c c c }
	\tablecaption{Initial Temperature Parameters.}
	\tablehead{\colhead{\parbox[t]{4em}{\centering Mass \\ ($M_{\oplus}$)}} & \colhead{$\epsilon_{m}$} & \colhead{\parbox[t]{4em}{\centering $\langle T_{m,i} \rangle$\\(K)}} &\colhead{\parbox[t][][t]{5em}{\centering $T_{c,i}$ \\(K)}}}
	\startdata
	1 & 1.19 & 3000 & 3810\\
	2 & 1.32 & 3330 & 4630\\
	3 & 1.44 & 3630 & 5350\\
	5 & 1.64 & 4130 & 6640
	\enddata
	\end{deluxetable}

\subsection{Initial parameters}

The initial temperature parameters for each of the super-Earth models are given in Table 5. Previous parameterized convection models have shown that initial temperatures do not significantly affect the thermal evolution beyond a few hundred Myr \citep[see e.g.][]{S80,MS89}. Therefore, the present models are all started with the same initial mantle potential temperature $T_{p}$, which is the temperature of the mantle adiabat extrapolated to the surface. The initial mantle potential temperature is set to 2520 K for all models, which is equivalent to an average mantle temperature of $\sim$ 3000 K for the Earth. The average mantle temperatures are calculated accordingly for all super-Earth models. In models described in a later section, which include core evolution, we assume an initial temperature contrast across the CMB of 100 K. Therefore, the initial core temperature is set equal to the temperature extrapolated along the mantle adiabat to the CMB plus 100 K.  We describe the effect of different initial temperatures in Section 4.

\section{Results}

For our nominal model, we focus on the evolution of temperature and water abundances in both the mantle and on the surface. In the following sub-section, we introduce a second model which includes core-cooling and uses a different characteristic mantle viscosity for the upper mantle that produces significantly different results. 

\subsection{Nominal Model - Single Layer Convection}

	 \begin{figure}
	\includegraphics[scale = 0.6,clip, trim = 0 0 0 0]{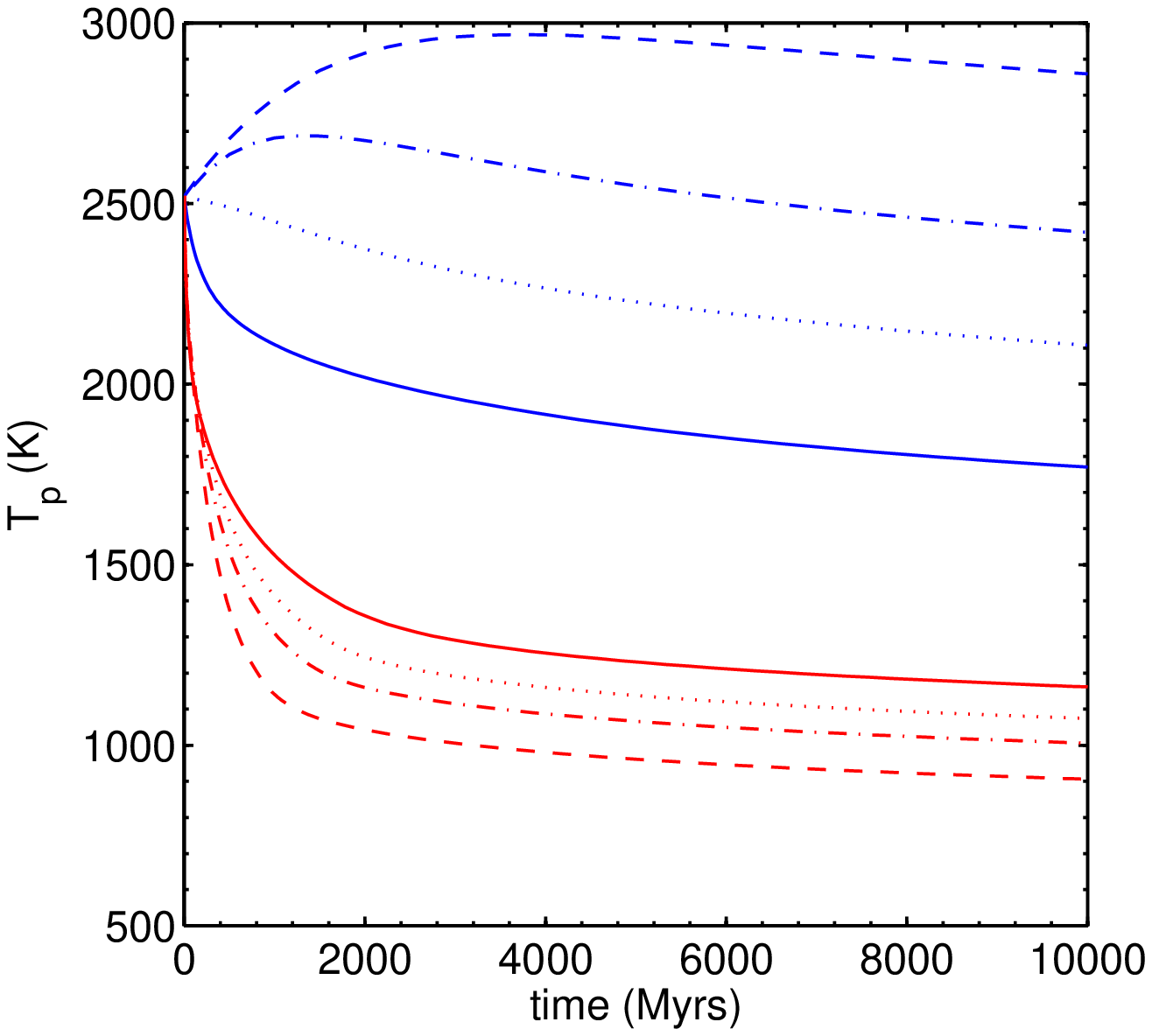}
	\includegraphics[scale = 0.6,clip, trim = -20 0 0 0]{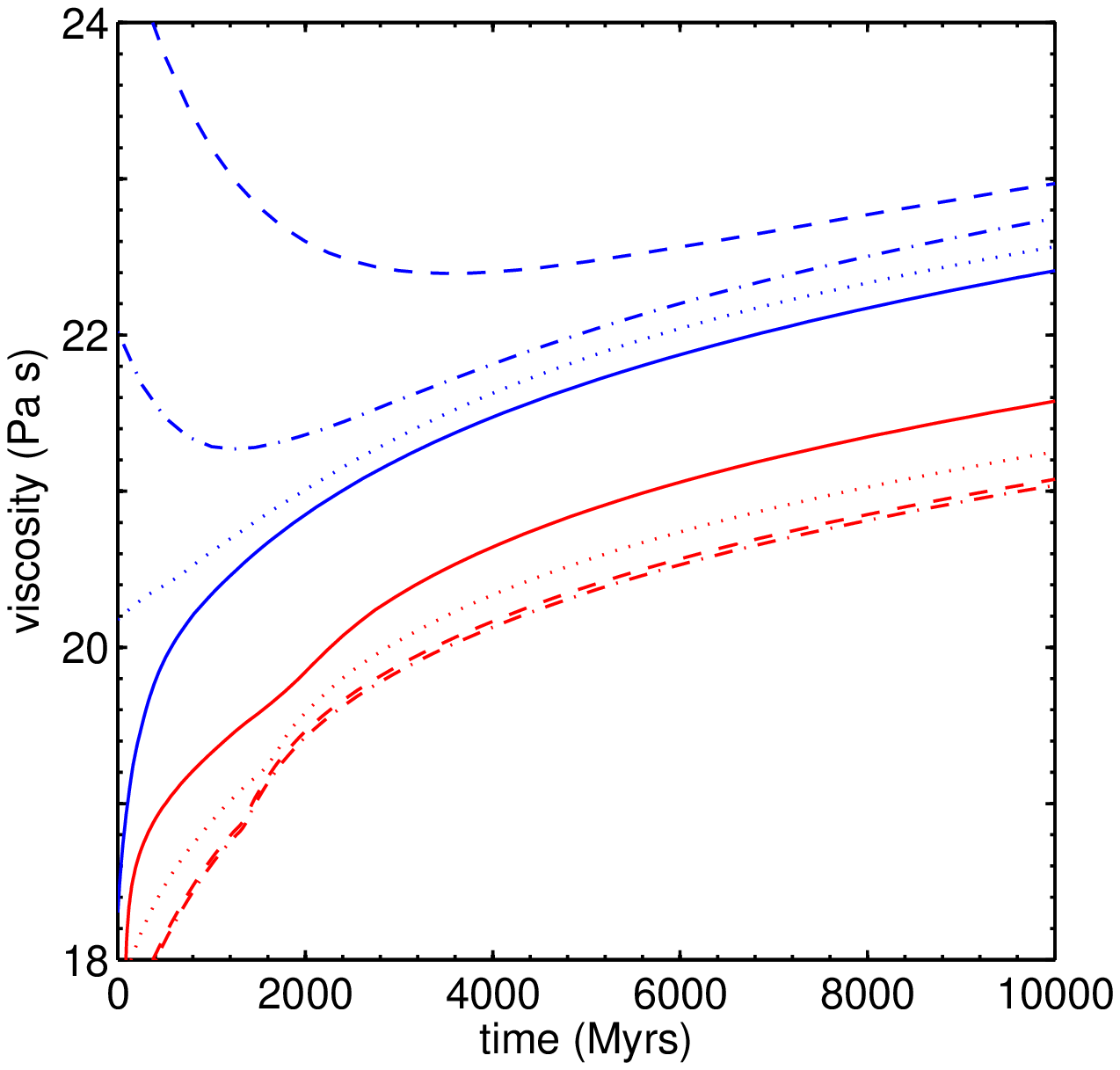}
	\includegraphics[scale = 0.58,clip, trim = -15 0 0 0]{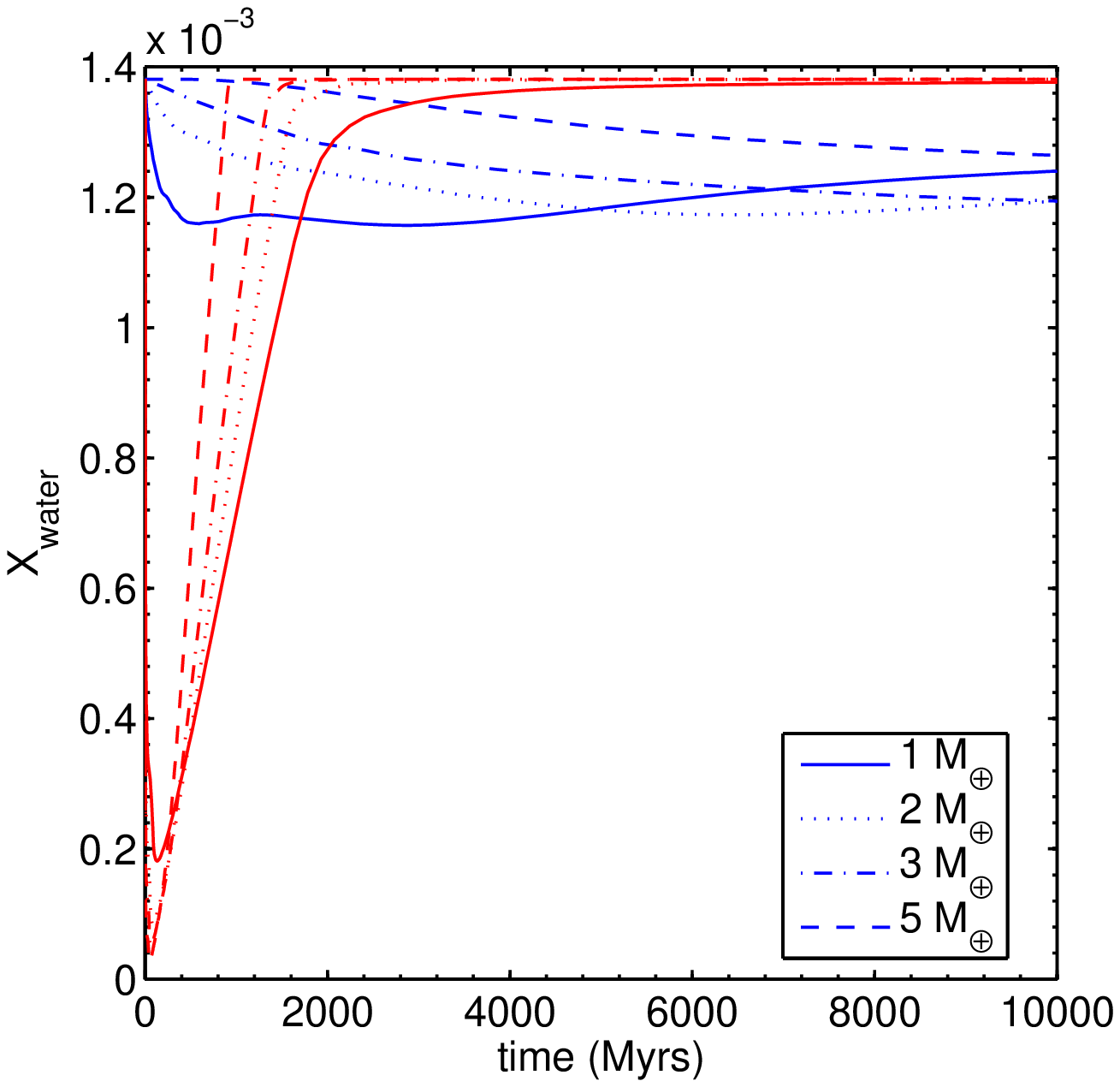}
	\caption{Nominal model with single layer convection. (a) mantle potential temperature, (b) mantle viscosity, and (c) mantle water mass fraction. Red lines represent calculations done with a pressure-independent viscosity, blue lines represent the use of a pressure-dependent viscosity. Line styles indicate planet mass according to the legend.}
	\end{figure}
	 \begin{figure}
	\plotone{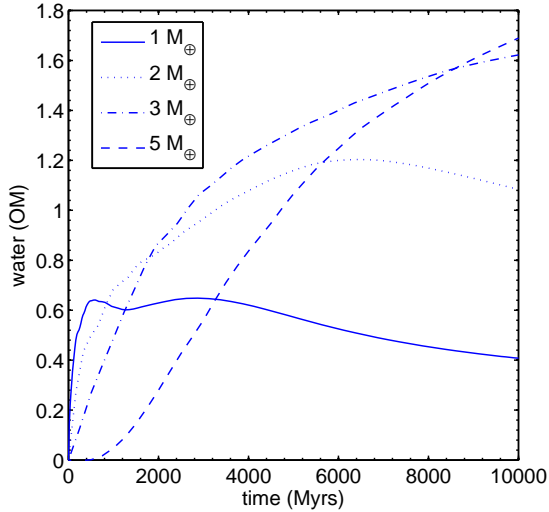}
	\caption{Surface water abundance for the nominal models. The water abundance is equivalent to 4 ocean masses of water in the Earth-sized planet.}
	\end{figure}
Figures 2 and 3 show results for the nominal model using parameters from Table 3 and a mantle water abundance of $\sim$1400 ppm (4 OM for 1 M$_{\oplus}$). Figure 2 shows the evolution of the mantle potential temperature, viscosity and water abundance for the nominal model using two different viscosity parameterizations. For the curves shown in red, the activation volume is set to 0, so the viscosity is pressure-independent. The blue curves include a non-zero activation volume (see Table 4), so the viscosity depends on pressure as well as temperature and water fugacity. The viscosity is calculated with the average mantle temperature $\langle T_{m}\rangle$ and the mid-mantle pressure. \\
\indent For the pressure-independent viscosity, Fig. 2a seems to indicate that the large planets cool off more rapidly than the smaller planets. The 5 $M_{\oplus}$ planet has a final potential temperature $\sim$240 K lower than that of the 1 $M_{\oplus}$ planet. This is counter to standard intuition: large planets should cool off more slowly than small planets due to their smaller surface-area/volume ratios. However, note that these are potential temperatures (i.e, the mantle temperature extrapolated to the surface along an adiabat). The average mantle temperature of the 5 M$_{\oplus}$ planet is in fact hotter than the 1 M$_{\oplus}$, but by only $\sim$80 K. The adiabatic gradient is much steeper for the large planets due to dependence on gravity, so the near-surface temperatures are therefore much lower. As seen in Fig. 2b, the hotter average mantle temperatures of the large planets results in lower mantle viscosities and therefore more rapid cooling. \\
\indent The lower near-surface temperatures of larger planets reduces the degree of melting, which strongly affects their water cycles, shown in Fig. 2c. There is a sharp decrease in mantle water abundance in the first 2 Gyr due to rapid early outgassing for all planets. As mantle temperature drops, near-surface temperatures drop below the solidus temperature, which halts melting and outgassing. This is followed by a rapid ingassing period. Ingassing is limited by the mass of the water at the surface and the thickness of the hydrated surface layer, which is regulated by the surface heat flux. All but the smallest planet have ingassed nearly all of their water by $\sim$2 Gyr. The deep water cycles of these planets have effectively ceased. However, it should be noted that for all planets, this leaves a small residual surface reservoir of water that effectively cannot be lost to the mantle. \\
\indent For the pressure-dependent viscosity calculations (blue curves) both the average and potential temperatures of the smaller planets cool more quickly than the larger planets (see Fig. 2a). In fact, the largest planets initially heat up substantially before beginning to cool, so near-surface melting persists for much longer than in the pressure-independent case. The final temperature of the 5 $M_{\oplus}$ planet is $\sim$1000 K higher than that of the 1 $M_{\oplus}$ planet. The slow cooling of the planets can be attributed to the substantially larger viscosities in this model, due primarily to the pressure dependence of the viscosity (see Fig. 2b). The changes in water abundance shown in Fig. 2c are much more gradual than with pressure-independent viscosities. There is a rapid early outgassing phase only for the smallest planet, which has the lowest mantle viscosity, but this outgassing is much less complete than for the pressure-independent case. The larger planets show steady but very gradual outgassing, as their temperatures increase and their viscosities drop. In fact, the 5 $M_{\oplus}$ planet has delayed onset of outgassing by $\sim$1 Gyr, due to a very large thermal boundary and low surface heat flux, both of which can be attributed to the large initial mantle viscosity. \\
\indent The activation volume that we use here for the olivine viscosity is 4 cm$^{3}$ mol$^{-1}$, which is derived from experimental data. However, the activation volume for other silicates has been shown to decrease with pressure \citep{Vlada11}. In their thermal models, \cite{Vlada12} use activation volumes for pressures at the planet\textquoteright s CMB (2.5 cm$^{-3}$ mol$^{-1}$ for Earth). Therefore, a slightly lower activation volume may be more appropriate, particularly for the larger planets. Intermediate values of $V_{a}$ give mantle temperatures intermediate to those shown in Fig. 2a for the pressure-dependent viscosities. The mantle water abundance drops less precipitously in early times, and ingassing is much slower than for the pressure-independent case, but more complete than for the pressure-dependent model shown in Fig. 2b. The pressure-dependence of the viscosity is therefore an important parameter to include in the models. The value of activation volume will affect the final volumes of the surface oceans and the mantle temperature.\\
\indent Figure 3 shows the surface abundance of water for the nominal model for each of the super-Earths. This figure is the surface corollary to Fig. 2c. We show only the pressure-dependent models for clarity. Note that while the relative abundance of water is the same for each planet, the total planetary water mass is 4 (8, 12, 20) ocean masses for the 1 (2, 3, 5 M$_{\oplus}$) planet. The time-dependent behaviors of the planets do not scale simply with planet mass. The 1 $M_{\oplus}$ planet has a much more significant outgassing phase at early times, followed by ingassing. The larger planets show much more gradual outgassing, with delayed onset of outgassing for the 5 $M_{\oplus}$ planet. Due to the delayed outgassing, the 5 $M_{\oplus}$ planet has less surface water than the 3 $M_{\oplus}$ planet for most of their lifetimes. However, the 3 and 5 $M_{\oplus}$ planets have roughly equal surface water abundances at 10 Gyr. For smaller values of the activation volume, the surface oceans will be substantially larger. \\

\subsection{Boundary Layer Convection}
Volatile evolution models have typically assumed single-layer convection, which has been shown to work well for the Earth. However, thermal evolution models that neglect volatile evolution often assume that mantle convection is controlled by boundary layer instability, in which convective instability is determined by the local conditions at the boundary layers rather than the mantle as a whole. We will now describe how these models differ from the nominal model described above. For the boundary layer models, we assume mixed heating, with a non-zero heat flux from the core (taken here to be both solid and isothermal) and a conductive lower boundary layer. 
Here we are less interested in reproducing the Earth than in exploring the behavior of the models for different planet masses. When using the same physically-plausible parameters here, we show that the two types of models give fundamentally different results, due to the choice of characteristic mantle viscosity. \\
For the boundary layer models, a second heat transfer equation is needed for the core:
	\begin{equation}
	\rho_c C_{p,c} V_c  \frac{dT_c}{dt}=-A_c q_c
	\end{equation}
where variables are as in eq. (1), but defined for the core rather than the mantle. The core is isothermal, and so is characterized by a single temperature $T_{c}$. We neglect radioactive decay in the core, as well as latent heat due to possible core freeze-out. The heat flux out of the core is given by:
	\begin{equation}
	q_c=k \frac{(T_c-T_l)}{\delta_c} 
	\end{equation}
where $\delta_{c}$ is the lower thermal boundary layer, $T_{c}$ is the core temperature, and $T_{l}$ is the mantle temperature at the top of the lower thermal boundary layer (see Fig. 1). We use a local Rayleigh number to define the thickness of the lower boundary layer. The boundary layer thickness is determined by setting it equal to its critical thickness, the point at which it becomes unstable to convection. 
	\begin{equation}
	\delta_c=\left(\frac{\kappa \eta(T_c,P_{cmb}) Ra_{crit,l}}{g \alpha \rho_m (T_c-T_l)}\right)^{\frac{1}{3}}
	\end{equation}
where $\eta(T_{c}, P_{cmb})$ is the viscosity of the lower boundary layer. For the lower boundary layer, we use a perovskite rheology. The choice of rheologies will be discussed more below. The critical Rayleigh number of the lower boundary layer is given by $Ra_{crit,l} = 0.28 Ra^{0.21}$ \citep{DS00}, where $Ra$ is the global Rayleigh number used in the definition of $\delta_{u}$. The global Rayleigh number is defined here as in eq. (6), except for the characteristic viscosity. For the characteristic viscosity, we use the value defined by the temperature and pressure at the base of the upper thermal boundary layer. This has a signifcant effect on the outcome of the models as shown below. \\
\indent The viscosity of olivine is used for the upper thermal boundary layer, but olivine is not a stable phase at the pressures and temperatures of the lower thermal boundaries. We therefore use the viscosity law for perovskite derived by \cite{Vlada11} for the lower thermal boundary. Parameter values are given in Table 4. The lower mantle of super-Earths likely consists of perovskite transitioning to post-perovskite for larger planets \citep[see e.g.][]{Val06,Tack13}. No experimental data exists on the water-dependence of the viscosity of perovskite, so the value of $r$ is set to zero. \cite{Vlada11} also give the dependence of the activation volume as a function of pressure. In the nominal models,  we use a value of 2.5 cm$^{3}$ mol$^{-1}$, which is the value at the Earth\textquoteright s CMB.\\
\indent We use a slightly smaller value of $\alpha$ for the lower thermal boundary layer (see Table 2), which is taken from \cite{Tack13} for the Earth\textquoteright s CMB. \cite{SB14} found that scaling $\alpha$ with planetary mass was as important as scaling $\rho_{m}$ to the calculation of planetary temperatures. However, we find little variability in the results for surface oceans when holding $\alpha$ constant. \\
\indent Results for the boundary layer convection model are discussed in the following subsection. Afterwards, we discuss how parameters affect the two different models.

\subsubsection{Results - Boundary Layer Convection Model}

	 \begin{figure*}
	\includegraphics[scale = 0.6,clip,trim = 0 0 0 0]{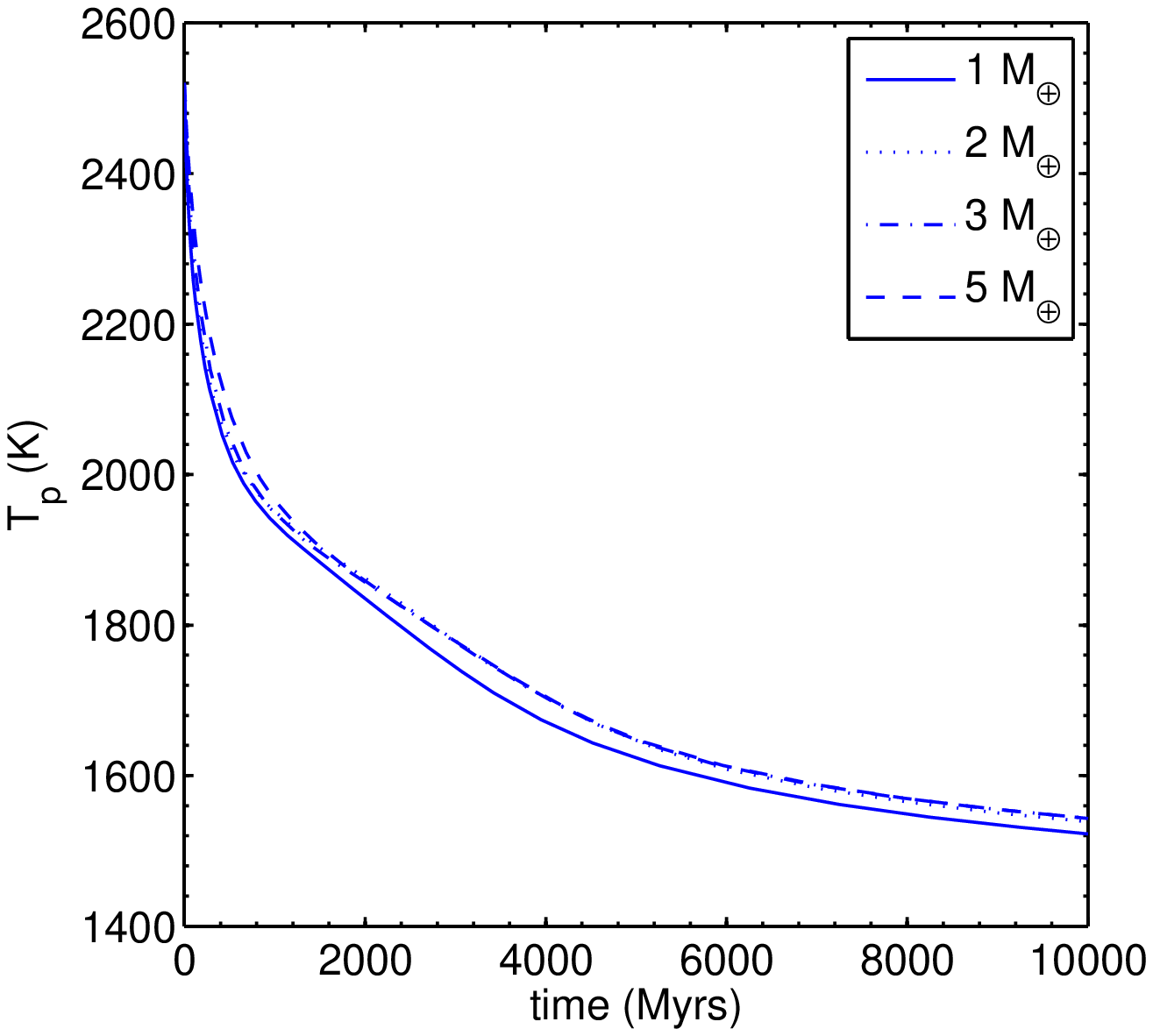}
	\includegraphics[scale = 0.6,clip, trim = 0 0 -40 0]{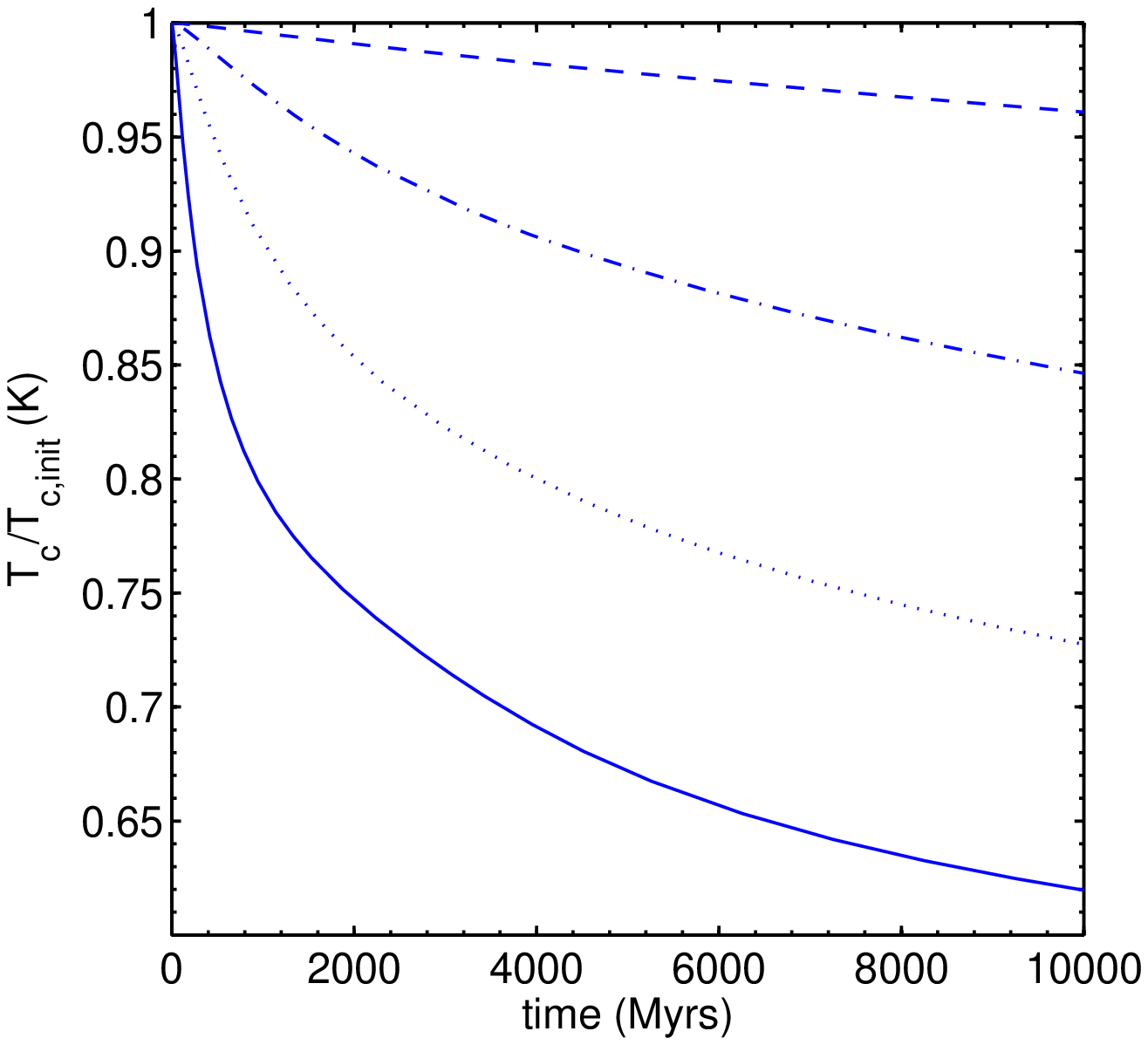}
	\includegraphics[scale = 0.58,clip, trim = -20 0 -20 0]{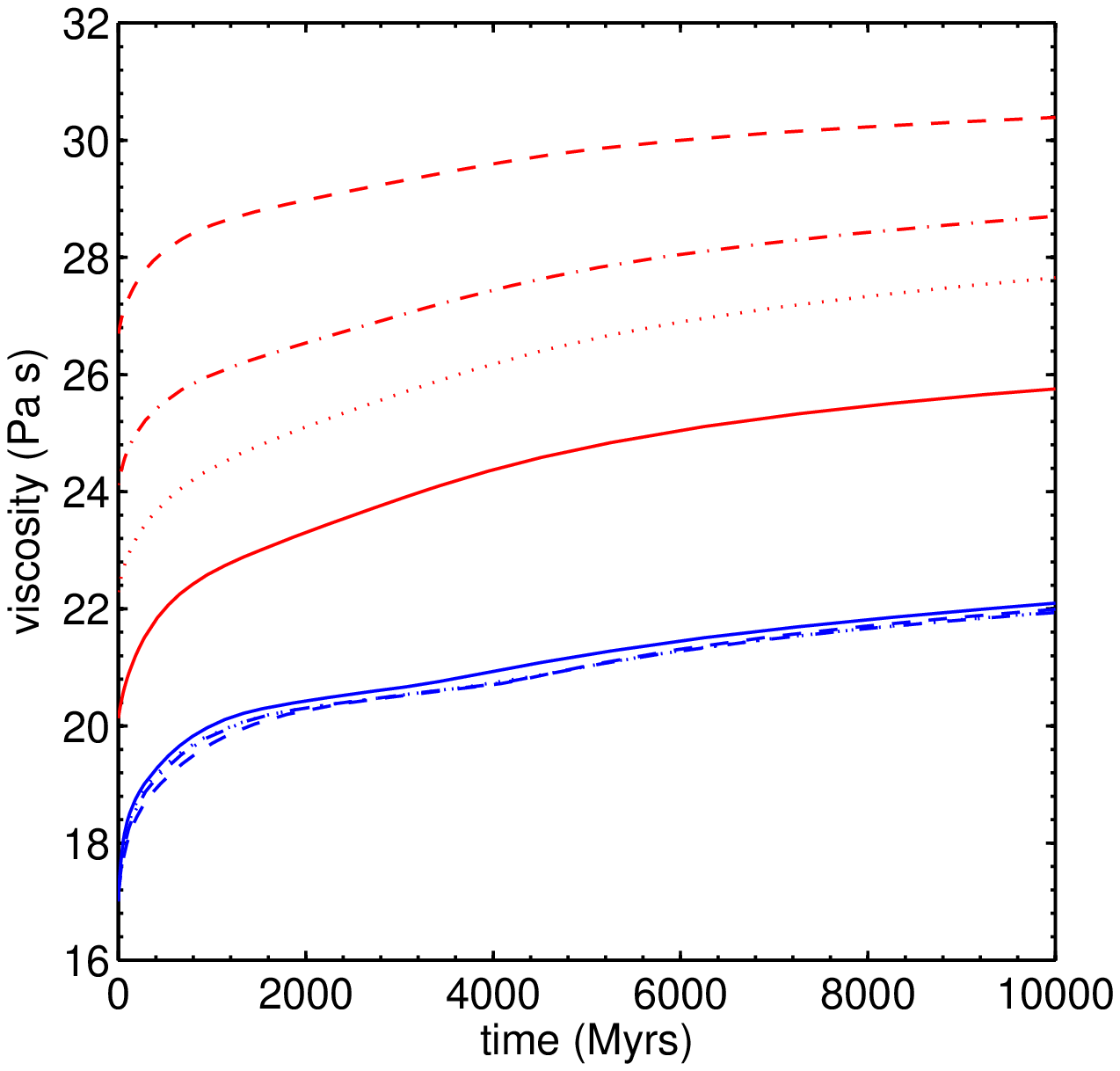}
	\includegraphics[scale = 0.58,clip, trim = -60 0 0 0]{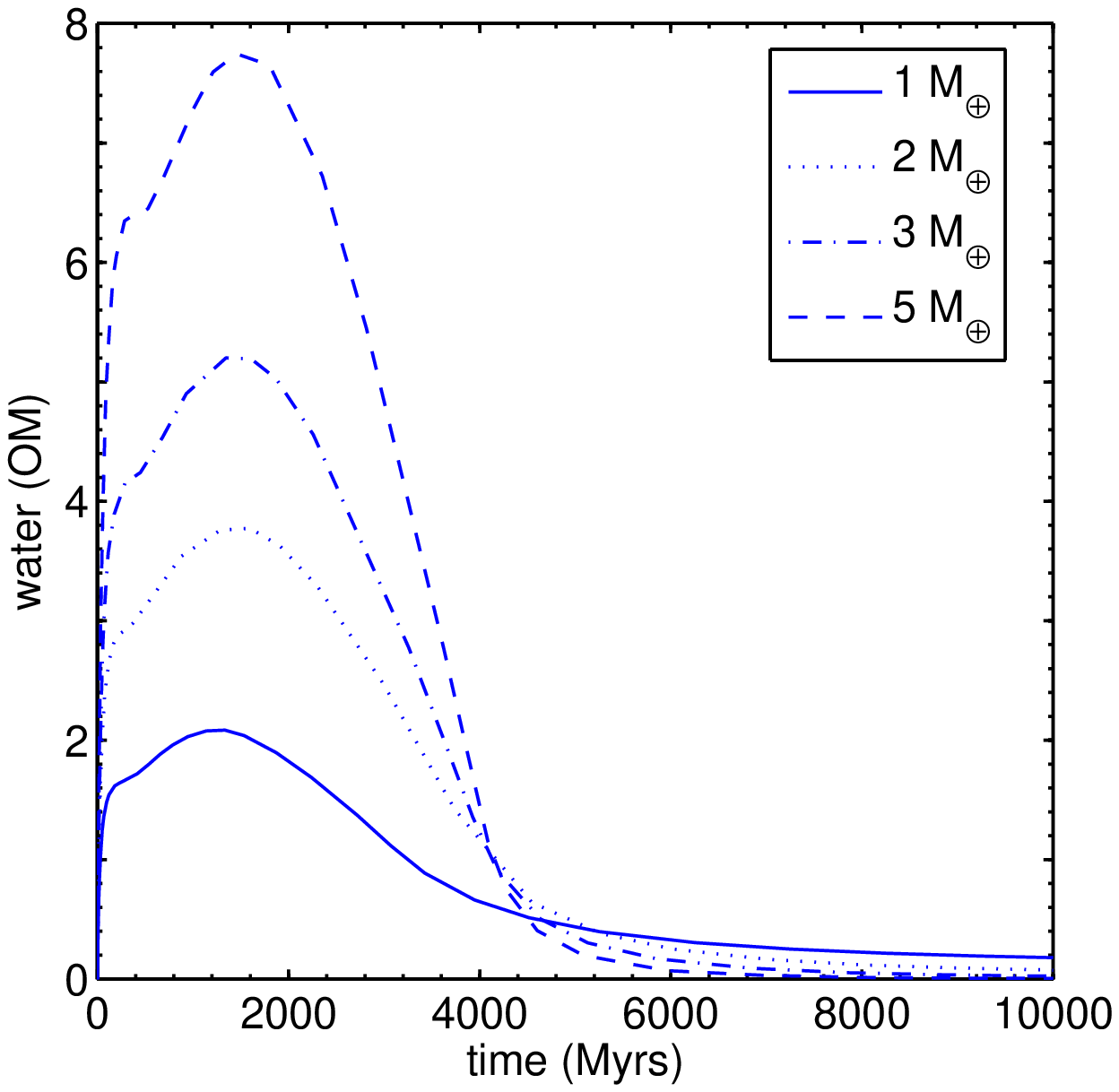}
	\caption{Results for nominal models with boundary layer convection heating. (a) mantle potential temperature, (b) core temperature normalized to initial core temperature, (c) thermal boundary layer viscosities (upper = blue, lower = red), (d) surface water abundance in ocean masses. Line styles indicate planet mass according to the legend.}
	\end{figure*}

Results are shown in Figure 4. Core temperatures are normalized to their initial values, because they vary by $\sim$3000 K. The upper boundary layer is calculated using the pressure-dependent olivine viscosity using temperature and pressure at the base of the boundary layer. The lower boundary is calculated using the pressure-dependent perovskite viscosity from \cite{Vlada11} using temperatures and pressures at the CMB. This viscosity does not depend on the mantle water abundance. \\
\indent Mantle potential temperatures decrease more rapidly and significantly here than for the nominal model. All planets have nearly identical potential temperature evolution, in contrast to the pressure-dependent results shown in Fig. 2a. The upper thermal boundary layer thickness is comparable for all planets, so the upper mantle viscosity used here is only weakly dependent on pressure (blue curves, Fig. 4c). In contrast, the viscosity of the lower thermal boundary layer is strongly dependent on pressure, and therefore the core temperature (Fig. 4b) is highly dependent on planet mass. The 1 M$_{\oplus}$ planet\textquoteright s core cools by $\sim$35\%, whereas the 5 $M_{\oplus}$ planet\textquoteright s core cools only $\sim$5\%. \\
\indent Figure 4d shows the surface water inventories for the boundary layer model, with 4 (8, 12, 20) OM of initial water. In comparison to Figure 3, it is obvious that significantly more water is outgassed from the mantle here. The 5 $M_{\oplus}$ planet has a peak surface water abundance of $\sim$8 OM, in comparison to $\sim$1.6 in Fig. 3. However, the residence time is much shorter. All of the planet\textquoteright s lose most of their surface water inventory back into the mantle by $\sim$4.5 Gyr. After complete ingassing, the 1 $M_{\oplus}$ planet has the largest remaining surface water abundance, with $\sim$0.2 OM at 10 Gyr. All planets have significantly lower surface water abundances at 10 Gyr for the boundary layer models than for the nominal models shown in Figure 3. For the boundary layer models, we find little difference between the pressure-dependent and pressure-independent viscosities, except for the cooling of the core. The peak surface water abundances remain essentially unchanged, but the presence of surface water does persist for $\sim$0.5 Gyr longer in the pressure-independent case. This is not surprising, since the pressures of the upper thermal boundary layer are very low and do not signficantly affect the viscosities. \\

\subsection{Dependence on parameters}

Many of the parameters used in the thermal and volatile evolution models are poorly constrained for the Earth, much less super-Earths. In the following section, we exam the effect of varying several of these parameters on the results of both the nominal and the boundary layer convection models. We refer the reader to \cite{SB14} for an analysis of the effect of $\beta$, $\alpha_{m}$, $\kappa_{m}$, and $\rho_{m}$ on thermal evolution models. 

	\begin{figure*}
	\plottwo{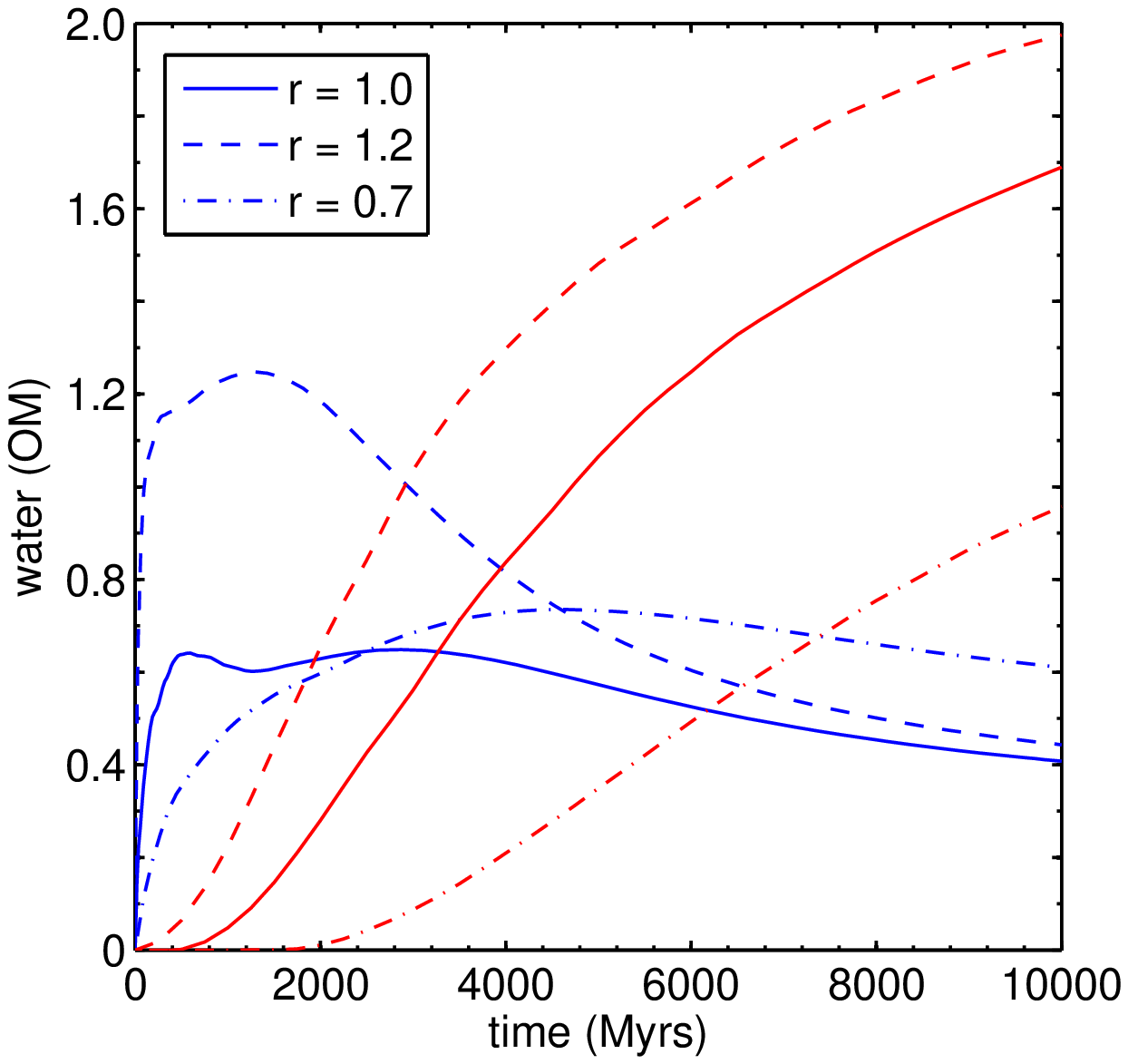}{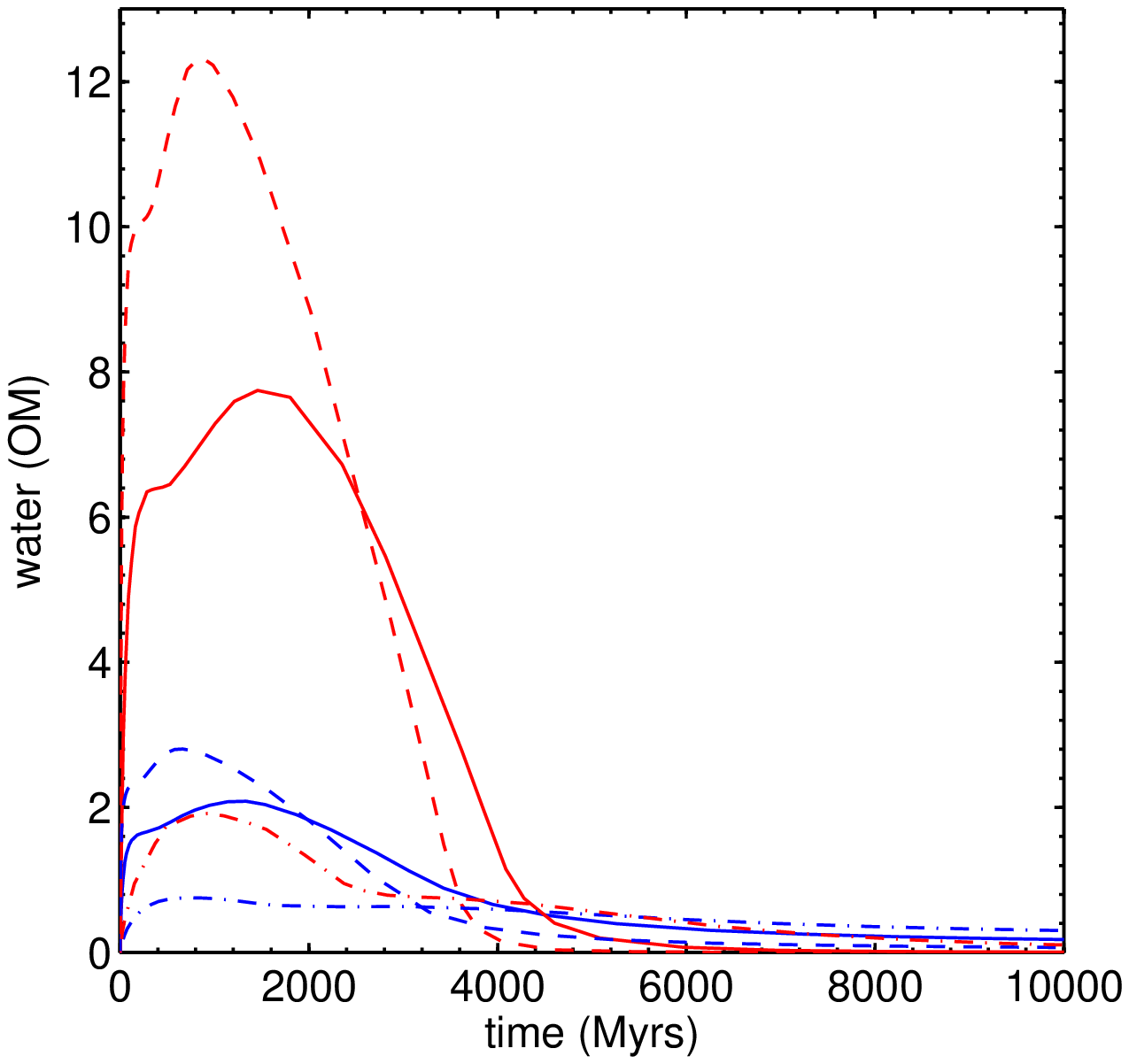}
	\caption{Surface water abundances for (a) the nominal single layer convection model and (b) boundary layer convection models using different values of the fugacity exponent $r$. Blue lines are for 1 $M_{\oplus}$, red lines for 5 $M_{\oplus}$.}
	\end{figure*}

	 \begin{figure*}
	\plottwo{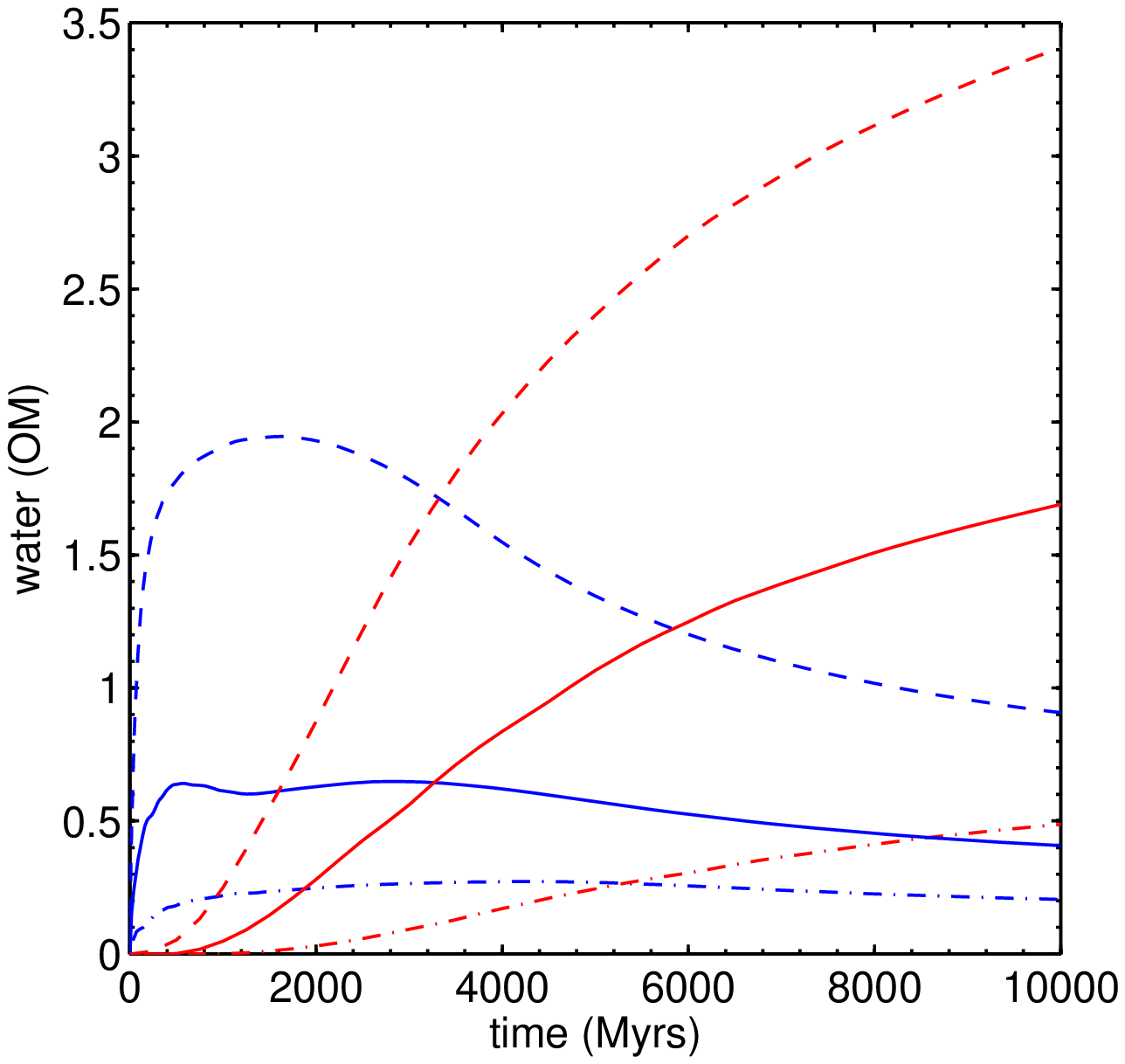}{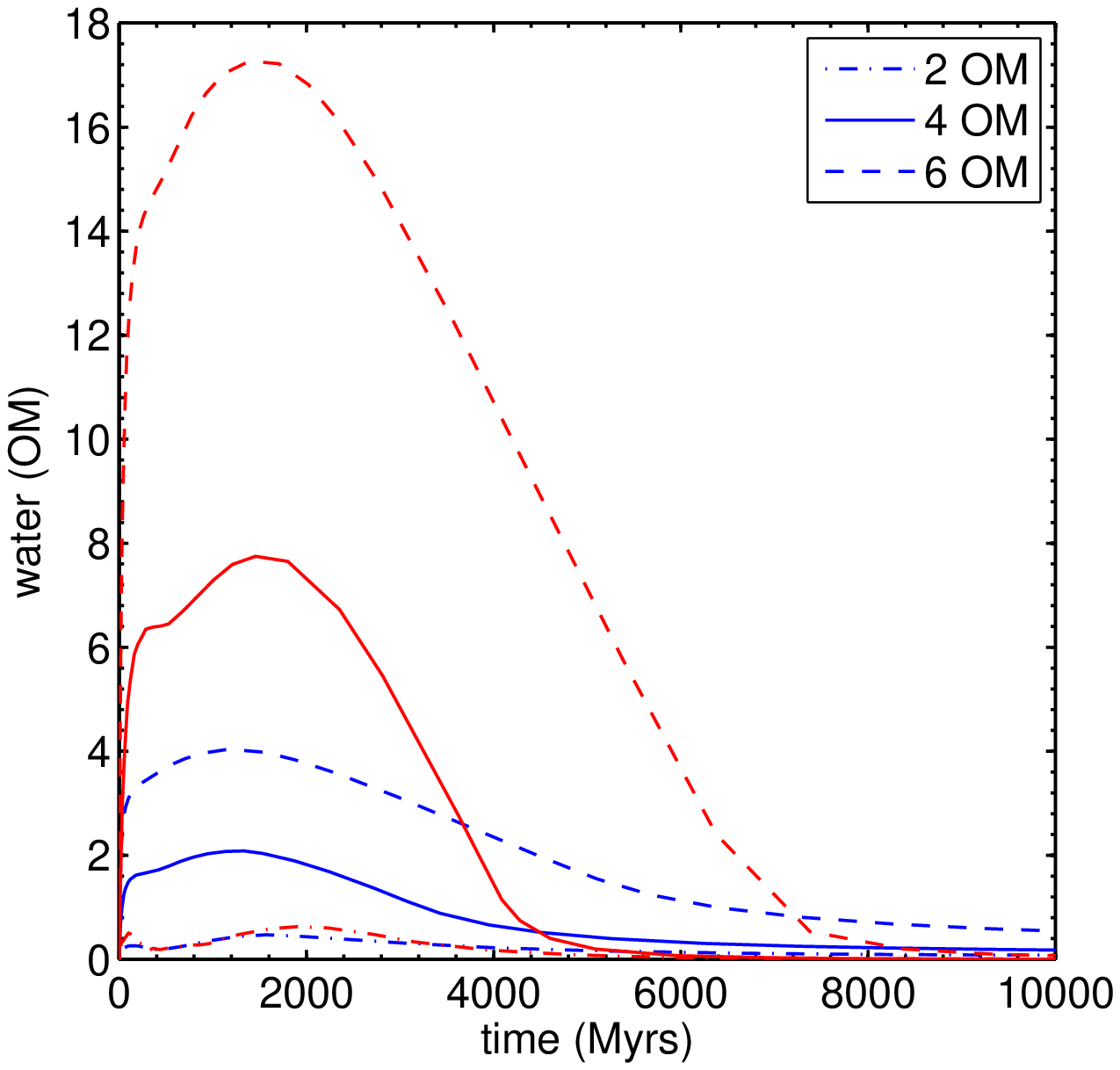}
	\caption{Surface water abundances for (a) the nominal single layer convection model and (b) boundary layer convection models for different initial water abundances. Total water abundances are equivalent to 2, 4 and 6 ocean masses of water for the Earth-mass planet. Abundances for the larger planets are the same in terms of mantle mass fraction of water. Colors indicate planet mass (blue 1 $M_{\oplus}$, red 5$M_{\oplus}$), and line styles indicate water abundance.}
	\end{figure*}

\subsubsection{Fugacity coefficient}

Figure 5 explores the effect of the fugacity coefficient $r$ on the results of the previous models for planets of 1 and 5 $M_{\oplus}$. Other planet masses are not shown for clarity. This figure shows $r$ values of 0.7, 1.0 (nominal), and 1.2. \cite{HK03} give values of $0.7 - 1.0$ for wet diffusion creep of olivine and $r = 1.2$ for wet dislocation creep of olivine, which is why we chose these values. It should be noted that the viscosities were not renormalized with the change of $r$ values. Therefore the variances reflect the absolute changes in viscosity. \\
\indent For the single layer convection model (Fig. 5a), the initial outgassing phase is significantly stronger for $r$=1.2 for the 1 M$_{\oplus}$ planet, but the final abundance is nearly the same as the nominal value. In contrast, outgassing is slightly delayed for $r$ = 0.7 but the final water abundance is $\sim$0.2 OM larger. For the 5 M$_{\oplus}$ planet, the larger $r$ value increases the amount of water outgassed, whereas the lower $r$ value both further delays outgassing and limits the total amount of water outgassed significantly.  \\
\indent For the boundary layer convection model (Fig. 5b), the higher $r$ value increases the amount of outgassing for both the 1 and 5 M$_{\oplus}$ planets and shifts the peak of outgassing to slightly earlier times. However, ingassing also occurs more rapidly, so the oceans persist only until $\sim$4 Gyr. The lower $r$ value reduces the amount of outgassing for both planets, and causes the surface water to persist for longer. For the 1 M$_{\oplus}$ planet and $r$ = 0.7, the water abundance is relatively constant over the planet\textquoteright s lifetime.  There is about 0.4 OM of water remaining on the surface after 10 Gyr. \\

\subsubsection{Total water abundance}

\indent Figure 6 shows the surface water abundances using different initial mantle water abundances. The surface water abundances appear to be a fairly straightforward function of water abundance. Models with larger water abundance have larger surface inventories. The outgassing occurs earlier for the single layer convection models (Fig. 6a), and oceans persist later for the boundary layer convection models (Fig. 6b). The effect of varying initial water abundances produces results similar to changes in the fugacity coefficient shown in Fig. 5. However, one major difference to note is that although both larger water abundances and larger fugacity coefficients increase outgassing, the persistence of the oceans differs. Surface oceans persist longer (til $\sim$8 Gyr for the 5 M$_{\oplus}$ planet) for enhanced water, whereas increasing the fugacity coefficient shortened the ocean lifetime. Another thing to note is that the surface water abundances for the boundary layer model with 2 (10) OM of water are nearly identical. \\

\subsubsection{Initial mantle temperature}

	\begin{figure}
	\plotone{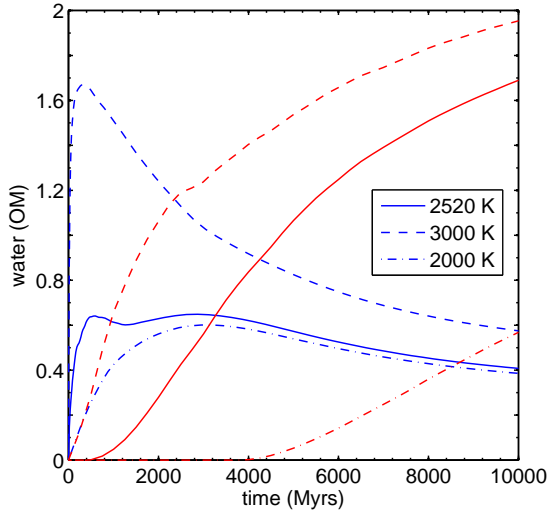}
	\caption{Surface water for the single layer convection model. Solid lines show the nominal model (also shown in Fig. 2 and 3), compared with models with either a higher (3000 K) or lower (2000 K) starting mantle potential temperature. Line styles indicate temperature according to the legend. Colors indicate planet mass (blue 1 $M_{\oplus}$, red 5$M_{\oplus}$).}
	\end{figure}

\indent Although models for the Earth show limited sensitivity to initial temperature (e.g. \cite{MS89}, \cite{TM92}), this appears not to be the case for the super-Earth models. Figure 7 compares results for the single layer model for the nominal starting mantle temperature of 2520 K, 2000 K, and 3000 K for the 1 (blue) and 5 (red) M$_{\oplus}$ planets. Mantle temperatures for the smaller planets converge within $\sim$ 4 Gyr on the nominal results (Fig 2a) but for the hotter starting temperature the 5 $M_{\oplus}$ planet remains persistently hotter throughout its lifetime. The hotter initial temperatures effect the water cycle for all of the planets. Although the 1 $M_{\oplus}$ planet converges to nearly the same temperatures, the initially hotter planet outgasses 3$\times$ more water within the first 500 Myr. The water is gradually ingassed over the planet\textquoteright s lifetime, but the final abundance of water remains slightly larger than for the colder starting planet. The 5 $M_{\oplus}$ planet begins outgassing much more rapidly, and continues steadily outgassing for its lifetime. It ends with $\sim$2 OM of water on the surface. Lower initial temperatures, which are more widely used in the literature \citep[e.g.][]{Vlada12,NB13}, delay and reduced the degree of outgassing, particularly for larger planets. For an initial potential temperature of 2000 K, the 5 $M_{\oplus}$ planet does not begin degassing until $\sim$4 Gyr. Given that surface water is likely necessary for plate tectonics \citep{K10b}, these planets may not experience plate tectonics at all, until very late in their lifetimes. These planets will likely start in a stagnant lid mode, which we have not attempted to model here. However, it is likely that the stagnant lid would allow the planet to heat up earlier so that formation of the oceans may not be as delayed as shown here. The boundary layer models show very limited dependence on the initial temperature and so are not shown here. For an initial starting temperature of 3000 K, the planets evolve at virtually the same temperatures, and the surface water abundances are only slightly enhanced. The different initial water abundances shown in Fig. 6 had a far larger impact on that model\textquoteright s results. \\

\subsubsection{Mid-ocean ridge length}

	\begin{figure}
	\plotone{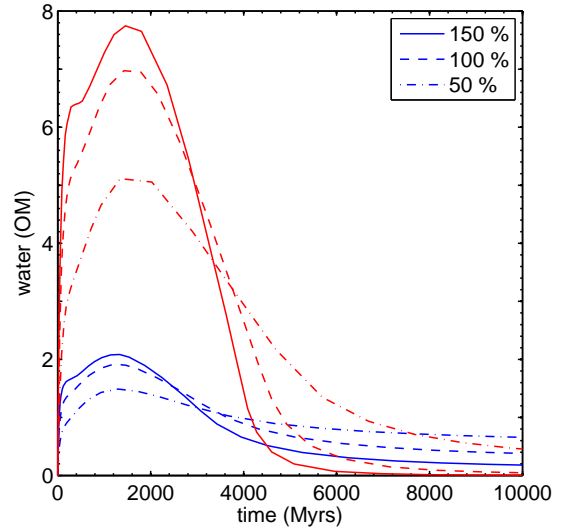}
	\caption{Surface water abundances for the boundary layer convection model. Solid lines show the nominal model (also shown in Fig. 4), compared with models using either  $L_{ridge}$ = 100\% (dash) or 50\% (dash-dot), respectively of the planetary circumference. Colors indicate planet mass (blue 1 $M_{\oplus}$, red 5$M_{\oplus}$). The nominal model uses $L_{ridge}$ = 150\% of the planetary circumference.}
	\end{figure}

\indent While we use a parameterization that produces the Earth's present day mid-ocean ridge length, we note that the length of the ridges on Earth may have changed throughout time. However, results for both models change only slighlty with different values for $L_{ridge}$. For the single layer model, the temperatures with variable $L_{ridge}$ do not change, and surface water abundances only slightly decrease. The effect on the water abundance for the boundary layer models is moderate, as shown in Figure 8. We show results for models using lower values of $L_{ridge}$, which would correspond with slower plate growth. The effect on temperature is minor for both the 1 and 5 $M_{\oplus}$ planet, so we do not show it here. For the surface water abundances, the lower $L_{ridge}$ values reduce the peak surface water abundance and significantly prolong the ingassing of the surface water after the initial outgassing phase. The 5 $M_{\oplus}$ planet has $\sim$0.5 OM of surface water remaining at 10 Gyr, whereas the 1 $M_{\oplus}$ planet has $\sim$0.8 OM of surface water remaining for an $L_{ridge}$ = 50\%.\\

\subsubsection{Other parameters}

\indent Another parameter that can significantly affect the model results is the viscosity parameterization. We have chosen here to normalize the viscosity to $10^{22}$ Pa s for a reference state of 1600 K, 0 Pa, and 500 ppm water. This gives a reasonable viscosity for the Earth\textquoteright s mantle. However, there are other choices that could be made. \cite{Sandu11} tuned their model to match the present day Earth\textquoteright s viscosity and heat flux, by normalizing their viscosity to $2.3\times 10^{21}$ Pa s at 2300 K and 500 ppm of water. Using this reference value for the mid-mantle pressure, we get $\eta_{0} = 2.5\times 10^{25}$ Pa s, roughly an order of magnitude lower than the value used here. For the single layer model, using this normalization factor results in mantle temperatures cooler by $\sim$100-150 K, and surface water abundances that resemble the hotter starting temperature in Fig. 7. \\
\indent The abundance of radioactive elements affects planets in both models. A recent paper shows that older planets, those formed early in the galaxy\textquoteright s lifetime, will have much lower abundances of radioactive elements, whereas younger planets may have up to 7 times more radioactive heat production  \citep{Frank14}. For the single layer models, increasing the uranium abundance by a factor of two increases the mantle temperatures by 100-200 K. In particular, all but the 1 $M_{\oplus}$ planet heat up within the first 2 Gyr, rather than cooling. The 5 $M_{\oplus}$ planet reaches a peak temperature of $\sim$3200 K, compared to $\sim$3000 K for the nominal results. The surface water abundances are increased slightly, but to a lesser extent than seen for a raise of initial temperature (Fig. 7). The mantle temperatures of the models heated from below increase by $\sim$100 K. The peak surface water abundances do not change from the nominal results, but the ingassing of the water back to the mantle takes a longer amount of time. The 1 $M_{\oplus}$ has substantial surface water until $\sim$ 4 Gyr, whereas the 5 $M_{\oplus}$ planet\textquoteright s surface water persists to $\sim$ 6 Gyr, compared to $\sim$4.5 Gyr for the nominal models. \\

\section{Discussion of Results}

\subsection{Stagnant lid regime}

Many terrestrial planets, such as Mars and Venus, do not experience plate tectonics, but are in a stagnant-lid regime. For these planets, a thick lithosphere insulates the convecting portion of the mantle from the surface. Stagnant lids develop when planets are too cool to maintain convection across the entire silicate layer. It is likely that most planets can transition between plate tectonics and the stagnant or sluggish lid regimes \citep{Sleep00}. However, the mechanisms by which this transition occurs are poorly understood. We can speculate, however, on how the transition would effect the models described. The formation of a stagnant lid insulates the mantle and allows the temperature to increase, which would lower the viscosity and increase convective vigor. Regassing would halt in this regime, as there is no transport mechanism for water to return to the mantle \citep[see e.g.][]{Morsch11}. For all of the planets in the boundary layer model, the formation of a stagnant lid would likely extend the lifetime of the surface oceans, by halting regassing. The planets could also heat up enough to re-initiate mantle melting and degassing. Future work will look at the effect of such a transition on the persistence of surface oceans. 

\subsection{Steady state versus thermal evolution}

As discussed in section 2.3, using the volatile evolution parameterization of \cite{MS89}, planets were found to achieve a volatile steady state. For a planet with a large global water abundance, most of the water will be found at the surface of the planet for the planet\textquoteright s lifetime. For planets with global water abundances smaller than the steady state value, all water will be trapped within the mantle. Note that \cite{CA14} considered a steady-state solution to determine the water mass fraction necessary for a planet to be considered a water planet (\textgreater 90\% surface ocean coverage). However, in the models considered here, steady state was never reached. In fact, the results discussed above show widely divergent outcomes for the same planet based upon the characteristic upper mantle viscosity chosen. Once the upper mantle temperature drops below the peridotite solidus, the melt layer disappears and the volatile cycle effectively ceases, although slow ingassing may continue until the surface becomes depleted of all water. \\

\subsection{Ocean depths on Super-Earths}

	\begin{figure*}
	\plottwo{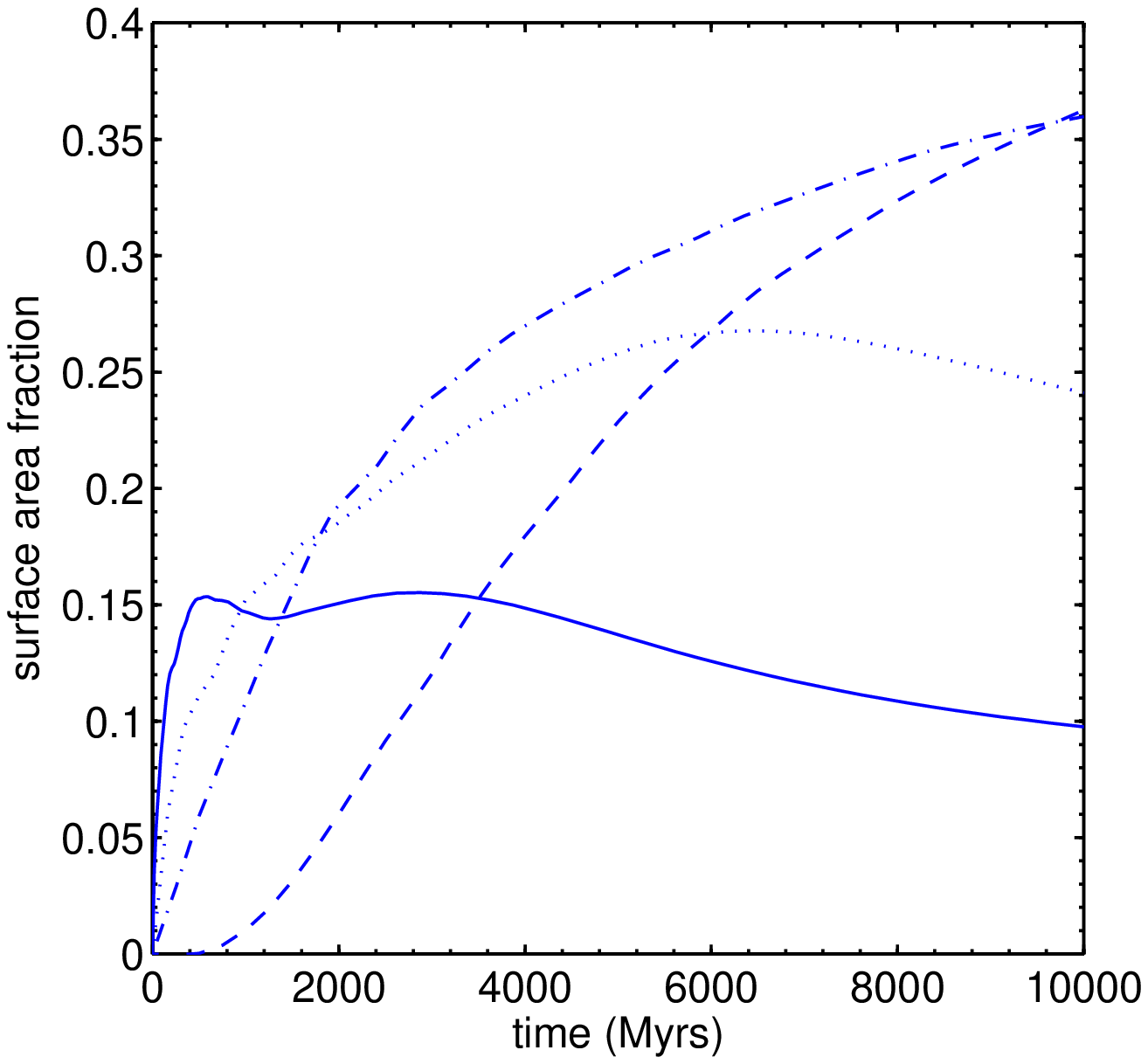}{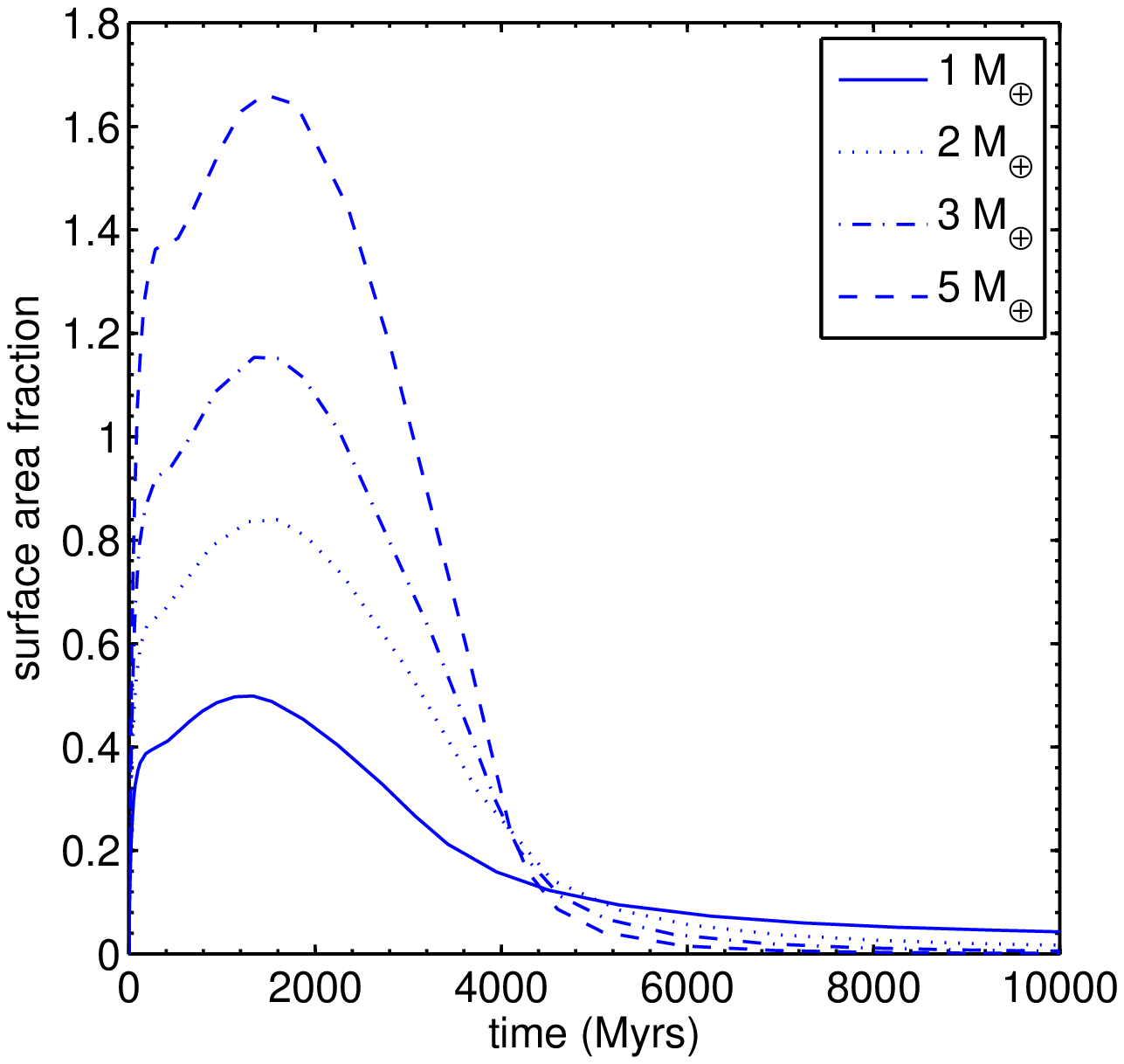}
	\caption{Minimum surface area coverage for the nominal models with a) single layer convection and b) boundary layer convection. The minimum surface area coverage is determined assuming the maximum ocean depth based on the scaling of \cite{CA14}. Planets with greater than 90\% ocean coverage are considered water planets.}
	\end{figure*}

We have focused in the discussion of results on the surface water abundances of the different models because this is a potentially observable parameter, and one that has implications for the possible habitability of super-Earths. Concerns have been raised about the habitability of, or more importantly the ability of life to begin on, planets with global ocean coverage. Some fraction of continental surface, which provides a shallow water environment, may be necessary for life to begin and for the evolution of complex life. \cite{CA14} derive a maximum ocean depth that separates planets with continents from totally water-covered planets using a crustal buoyancy model. They find that maximum ocean depth scales with surface gravity as $d_{max} \sim 11.4 \left(g / g_{\oplus} \right)^{-1}$ km. Applying this to the results for the models above gives us the minimum surface area coverage on each planet. We show these results for both the single layer and boundary layer models in Figure 9. The single layer convection planets have slightly less than 50\% areal coverage by oceans. The 5 $M_{\oplus}$ planet has minimal ocean areal fraction until $\sim$2 Gyr, which would suggest that life would be difficult to begin on such a planet in its infancy. For boundary layer convection, the 3 and 5 M$_{\oplus}$ planets have areal coverages greater than 1, which indicates that they will have global oceans. These planets are not likely to have exposed continents. However, the 1 and 2 M$_{\oplus}$ planets have significant continental area for both convection modes, which indicates that these planets may be best places to search for life. \\
\subsection{Additional planetary processes}

The models presented here are not comprehensive parameterizations of processes that can affect the water budget of a planet's surface. One particularly important process is the loss of water from a planet\textquoteright s atmosphere \citep{WP13}. Our model is agnostic to the form that water takes at the surface (i.e., water, ice, steam, etc.), other than by constraining the surface temperature. However, the surface temperature will vary with stellar type, orbital distance, atmospheric composition, and age. As \cite{WP13} show, loss of water vapor depends not only on the surface temperature, but also on the CO$_{2}$ abundance of the atmosphere. Significant loss of water vapor on hot CO$_{2}$-rich planets would reduce the amount of regassing, so the mantle would become more dehydrated over time. We also neglect continental crust formation and weathering \citep{Honing14}. Continental crust effectively acts as an insulating barrier, but also serves as a sink for radioactive elements which are extracted from the mantle. Weathering of continents and sedimentation rates, which are enhanced by living organisms, can affect the regassing rate into the oceans. \cite{Honing14} suggest that planets without life will evolve to have more surface water (therefore lower continental coverage), and dryer mantles than planets with life. Weathering also acts as climate control by stabilizing atmospheric CO$_{2}$, which affects the retention of water in the atmosphere as described above \citep{Abbot12}. Much further work needs to be done to truly understand the feedbacks that will contribute to the persistence of oceans on super-Earths. 

\section{Summary}

We explored two different scaling parameterizations for plate tectonics planets using either single layer convection or boundary layer convection. Mantle temperatures are significantly hotter for the first model, and the surface water abundance lower, but more persistent. For many different parameters, smaller planets will have initially larger surface oceans, but lose them more rapidly than larger planets. Larger planets show delayed outgassing, which may compromise them as locations on which life can originate. The boundary layer convection model cools very rapidly due to vigorous convection driven by both an upper and a lower thermal boundary layer. The surface water abundances on the massive planets in the boundary layer model are extremely high and suggest that these planets will have limited, if any, continental coverage. However, these oceans persist for less than half of the planet\textquoteright s lifetime. Upon complete ingassing of the oceans, the massive planets are effectively tectonically dead, and therefore unlikely to be habitable.\\
\indent Observations of rocky exoplanets in the 1 - 5 M$_{\oplus}$ range are already producing very accurate planet radii and mass determinations \citep{Dressing14}. Many of these exoplanets have ages, determined from asteroseismic ages of their host stars. In the era of JWST and large ground-based telescopes, some of the nearest exoplanets will have atmospheric spectroscopy capable of distinguishing different geochemical regimes, like some of the extremes described here. Understanding the general features of the deep water cycle across rocky planets of different mass and structure might give us unique windows into their interior through its singular effect on atmospheric and surface water abundance. \\

\acknowledgments
We thank Li Zeng for helpful discussions and an anonymous referee for a detailed review that greatly enhanced the quality of this paper.

\end{document}